\documentclass[twocolumn]{aastex631}
\usepackage{graphics,graphicx}
\hypersetup{urlcolor=blue}
\usepackage{natbib}
\usepackage[outdir=./]{epstopdf}
\usepackage{textcomp,gensymb}
\usepackage{xfrac}
\usepackage{multirow}

\usepackage{enumitem}

\newcommand{\teff}{\ensuremath{T_{\text{eff}}}}

\newcommand{\starname}{HIP 67522}
\newcommand{\planetname}{HIP 67522\,c}
\newcommand{\planetb}{HIP 67522\,b}

\newcommand{\re}{$R_{\oplus}$}
\newcommand{\gaia}{\textit{Gaia}}

\usepackage[outdir=./]{epstopdf}
\usepackage{ulem}
\usepackage{mathrsfs}
\usepackage{fontawesome}
\usepackage{comment}
\usepackage[version=4]{mhchem} 

\newcommand{\tess}{\textit{TESS}}
\newcommand{\ktwo}{{\textit K2}}

\newcommand{\jwst}{\textit{JWST}}
\newcommand{\notch}{\texttt{N\&L}}

\newcommand{\newedit}[1]{#1}

\newcommand{\unc}{Department of Physics and Astronomy, The University of North Carolina at Chapel Hill, Chapel Hill, NC 27599, USA}
\newcommand{\abc}{Astrobiology Center, 2-21-1 Osawa, Mitaka, Tokyo 181-8588, Japan}
\newcommand{\naoj}{National Astronomical Observatory of Japan, 2-21-1 Osawa, Mitaka, Tokyo 181-8588, Japan}

\newcommand{\sokendai}{Department of Astronomy, School of Science, The Graduate University for Advanced Studies (SOKENDAI), 2-21-1 Osawa, Mitaka, Tokyo, Japan}

\newcommand{\iac}{Instituto de Astrof\'\i sica de Canarias (IAC), 38205 La Laguna, Tenerife, Spain}

\newcommand{\komabasc}{Komaba Institute for Science, The University of Tokyo, 3-8-1 Komaba, Meguro, Tokyo 153-8902, Japan}

\pdfoutput=1

\shorttitle{} 
\shortauthors{}

\begin{document}

\title{TESS Investigation - Demographics of Young Exoplanets (TI-DYE) II:\\ a second giant planet in the 17-Myr system HIP 67522}

\correspondingauthor{Madyson G. Barber}
\email{madysonb@live.unc.edu}  

\author[0000-0002-8399-472X]{Madyson G. Barber}
\altaffiliation{NSF Graduate Research Fellow}
\affiliation{\unc} 

\author[0000-0001-5729-6576]{Pa Chia Thao}
\altaffiliation{NSF Graduate Research Fellow}
\altaffiliation{Jack Kent Cooke Foundation Graduate Scholar}
\affiliation{\unc} 

\author[0000-0003-3654-1602]{Andrew W. Mann}
\affiliation{\unc}

\author[0000-0001-7246-5438]{Andrew Vanderburg}
\affiliation{Department of Physics and Kavli Institute for Astrophysics and Space Research, Massachusetts Institute of Technology, Cambridge, MA 02139, USA}

\author[0000-0003-1368-6593]{Mayuko Mori}
\affiliation{\abc}
\affiliation{\naoj}

\author[0000-0002-4881-3620]{John H. Livingston}
\affiliation{\abc}
\affiliation{\naoj}
\affiliation{\sokendai}

\author[0000-0002-4909-5763]{Akihiko Fukui}
\affiliation{\komabasc}
\affiliation{\iac}

\author[0000-0001-8511-2981]{Norio Narita}
\affiliation{\komabasc}
\affiliation{\iac}
\affiliation{\abc}

\author[0000-0001-9811-568X]{Adam L. Kraus}
\affiliation{Department of Astronomy, The University of Texas at Austin, Austin, TX 78712, USA}

\author[0000-0003-2053-07492]{Benjamin M. Tofflemire}
\altaffiliation{51 Pegasi b Fellow}
\affiliation{Department of Astronomy, The University of Texas at Austin, Austin, TX 78712, USA}

\author[0000-0003-4150-841X]{Elisabeth R. Newton}%
\affiliation{Department of Physics and Astronomy, Dartmouth College, Hanover, NH 03755, USA}


\author[0000-0002-4265-047X]{Joshua~N.~Winn}
\affiliation{Department of Astrophysical Sciences, Princeton University, 4 Ivy Lane, Princeton, NJ 08544, USA}

\author[0000-0002-4715-9460]{Jon M. Jenkins}%
\affiliation{NASA Ames Research Center, Moffett Field, CA, 94035, USA}

\author[0000-0002-6892-6948]{Sara~Seager}%
\affiliation{Department of Physics and Kavli Institute for Astrophysics and Space Research, Massachusetts Institute of Technology, Cambridge, MA 02139, USA}
\affiliation{Department of Earth, Atmospheric and Planetary Sciences, Massachusetts Institute of Technology, Cambridge, MA 02139, USA}
\affiliation{Department of Aeronautics and Astronautics, MIT, 77 Massachusetts Avenue, Cambridge, MA 02139, USA}


\author[0000-0001-6588-9574]{Karen A.\ Collins}
\affiliation{Center for Astrophysics \textbar \ Harvard \& Smithsonian, 60 Garden Street, Cambridge, MA 02138, USA}

\author[0000-0002-6778-7552]{Joseph D.\ Twicken}
\affiliation{SETI Institute, Mountain View, CA 94043 USA}
\affiliation{NASA Ames Research Center, Moffett Field, CA, 94035, USA}

\begin{abstract}
The youngest ($<$50\,Myr) planets are vital to understand planet formation and early evolution. The 17\,Myr system \starname\ is already known to host a giant ($\simeq$10$R_\oplus$) planet on a tight orbit. In the discovery paper, Rizzuto et al. 2020 reported a tentative single transit detection of an additional planet in the system using \tess. Here, we report the discovery of HIP 67522 c\newedit{, a 7.9 \re\ planet} which matches with that single transit event. We confirm the signal with ground-based multi-wavelength photometry from Sinistro and MuSCAT4. At a period of 14.33 days, planet c is close to a 2:1 mean motion resonance with b (6.96 days or 2.06:1). The light curve shows distortions during many of the transits, which are consistent with spot crossing events and/or flares. Fewer stellar activity events are seen in the transits of planet b, suggesting that planet c is crossing a more active latitude. Such distortions, combined with systematics in the \tess\ light curve extraction, likely explain why planet c was previously missed.

\end{abstract}


\section{Introduction} \label{sec:intro}
Young ($<500$\,Myr) planets provide an opportunity to detect planet-sculpting processes as they happen. In particular, young multi-planet systems are valuable because they allow for better control of variables within a single system and enable a wider range of science cases than those with single planets. Such systems enable tests on the origin of intrasystem uniformity \citep[e.g.,][]{Lissauer2011, Lammers2023}, \newedit{differential measurements of planetary atmsopheres within a system \citep[comparative planetology, e.g.,][]{Barat2024}}, and \newedit{searches} for transit timing variations that may yield precise masses and eccentricities \citep[e.g.][]{Steffen2012,Masuda2014}, among a wide range of other science cases. 

Stellar, and hence planetary, ages are generally challenging to determine for individual systems \citep{Soderblom2014}. Therefore, planets in stellar associations are critical in understanding planet formation mechanisms, as the planets' ages can be precisely derived from the parent stellar population. Studies from the small sample of young planets ($< $30) have demonstrated several key findings: some close-in planets either form in situ or migrate quickly \citep{David2019b, THYMEVI, Wood2023}, young planets tend to be larger than their older counterparts \citep{Fernandes2022, Fernandes2023, Vach2024}, and \newedit{among the six $<100$\,Myr planets with Rossiter–McLaughlin or Doppler tomography, all are aligned or nearly aligned \citep[e.g.][]{Hirano2020, Johnson2022, Hirano2024}.}

\newedit{\tess\ \citep{Ricker2015} and \ktwo\ \citep{Howell2014} have played critical roles in the discovery and characterization of young planetary systems. Light curves from these missions have} been used to discover systems as young as 11\,Myr \citep{THYMEVI, Zakhozhay2022} and young planets as small as $\simeq1R_\oplus$ \citep{Livingston2018,Capistrant2024}, as well as aided in the discovery of new young stellar associations hosting transiting planets \citep[e.g,][]{THYMEV, Nardiello2022, Thao2024}.

One such important discovery is the young, hot, $10~R{_\oplus}$ \planetb\ \citep{THYMEII}. The planet was identified using the \texttt{Notch} \& \texttt{LoCoR} pipeline \citep[\notch;][]{Rizzuto2017} and \tess\ photometry. At 17\,Myr, \planetb\ stands as one of the youngest transiting planets known. It orbits a bright ($V=9.8$) G-star in the nearby Scorpius-Centaurus OB association, which made it a prime target for transmission spectroscopy with \jwst\ \citep{MannJWSTProp}, Rossiter-McLaughlin observations \citep{Heitzmann2021}, and studies of atmospheric escape (Milburn et al. in prep). 

\citet{THYMEII} noted a single trapezoid-like event in the light curve, consistent with an additional transiting planet (\planetname). However, they detected no additional transit-like signals and hence could not confirm the signal as planetary in origin. Lacking evidence of another transit, it was assumed the candidate planet had a period $>$24 days (based on the time from the transit to the end of the sector). Based on the observed transit duration, the estimated orbital period was between 30 and 124 days (with 68\% confidence). 

Since the discovery, \tess\, re-observed the \starname\ system in two more sectors, and many teams have made improvements in the handling of \tess\ light curves for young stars \citep[e.g.][]{Capistrant2024}. Using this data and our updated pipeline, we recover the original candidate transit alongside four other consistent transit-like events, yielding a period of 14.33\,days (henceforth \planetname). This places \planetname\ close to the 2:1 resonance with \planetb. The planet's large size, proximity to a mean-motion resonance, and age make it a compelling target for additional follow-up.

\section{\tess\ light curve} \label{sec:tess}
\starname\ (TIC 166527623; TOI-6551; HD 120411) was first observed by \tess\ in Sector 11, from 2019 April 23 to 2019 May 20, and was re-observed in Sector 38 (2021 April 29 - 2021 May 26) and Sector 64 (2023 April 6 - 2023 May 3). The target was pre-selected for 2-minute short-cadence light curves for Sector 11 (G011280, PI A. Rizzuto) and Sector 38 (G03141, PI E. Newton and G03130, PI A. Mann) and 20-second light curves for Sector 64 (G05015, PI B. Hord and G05106, PI E. Gillen). In total, \tess\ observed five transits of \planetname.

\subsection{Extraction pipeline}
For our analysis, we used a custom light curve extraction pipeline starting from Simple Aperture Photometry \citep[SAP;][]{Twicken2010SAP, Morris2020SPOC} \newedit{fluxes from} the Science Processing Operations Center \citep[SPOC;][]{Jenkins2016SPOC}. We used the shortest cadence data available for each sector. All \tess\ light curves used in this analysis can be found in MAST: \dataset[10.17909/9deq-2151]{http://dx.doi.org/10.17909/9deq-2151} 

We applied systematic corrections following the prescription in \citet{Vanderburg2019}. To summarize, we corrected the SPOC SAP light curves with a linear model consisting of a basis spline (B-spline) with regularly-spaced breaks at 0.2 day intervals to model long-term, low-frequency stellar variability, several moments of the distribution of the spacecraft quaternion time series measurements within each exposure, seven co-trending basis vectors from the SPOC Presearch Data Conditioning \citep[PDC;][]{Stumpe2012, Smith2012, Stumpe2014} band-3 flux time series correction with the largest eigenvalues, and a high-pass (0.1 day) time series from the SPOC background aperture. \newedit{Additional details required to reproduce this correction can be found in \cite{Vanderburg2019}.}

We estimated the uncertainties on the flux by taking the median value of three different methods: 1) the median absolute deviation of the point minus the adjacent point; 2) the mean absolute deviation of the flattened light curve \citep[flattened using \texttt{lightkurve};][with a 4$\sigma$ outlier threshold]{lightkurve}; and 3) sigma clipping, applying a median filter, sigma clipping again, and fitting a Gaussian to the resulting distribution of points. Each method was broadly consistent with each other, and the final results do not depend on the uncertainty estimate. We found the uncertainties for each sector separately and adopted an uncertainty of 0.0007, 0.0009, and 0.0014 for Sectors 11, 38, and 64, respectively. 

\begin{figure*}
    \centering
    \includegraphics[width=\linewidth]{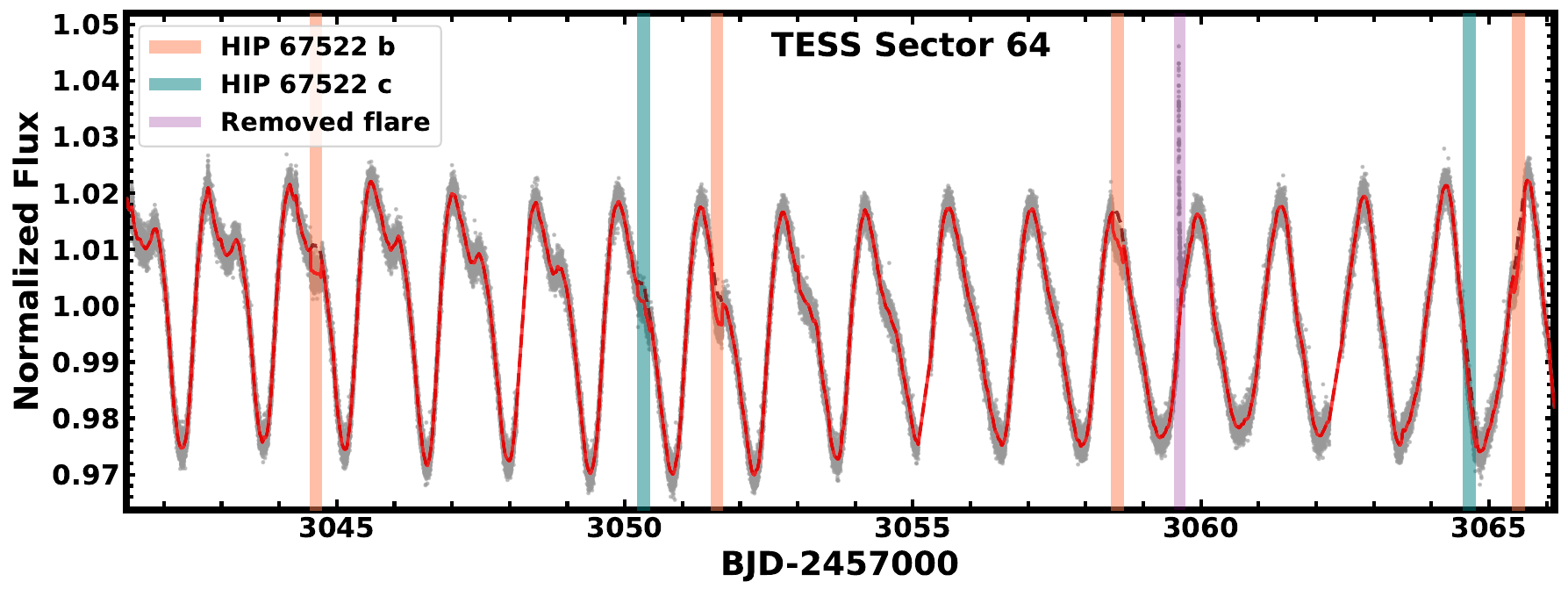}
    \caption{Representative section of the \tess\ light curve (gray points) with the best-fit GP model (red line). The locations of transits of \starname\ b and c \newedit{and the location of the removed flare} are shown as the highlighted regions.}
    \label{fig:gpsector}
\end{figure*}

\section{Identification of the transit} \label{sec:transit}

As part of our search for transiting planets in the youngest stellar associations (Barber et al. submitted), and motivated by both the single-transit detection in \cite{THYMEII} and the recent transit timing variations (TTVs) detected in \planetb\ (Thao et al. submitted), we searched \starname\ for additional planet signals. We used the updated \notch\ pipeline \citep{Rizzuto2017}\footnote{\newedit{\url{https://github.com/arizzuto/Notch_and_LOCoR}}} as described in Barber et al. (submitted). To summarize, we updated the default box-least squares (BLS) search to the more recent implementation in \texttt{astropy}\footnote{http://www.astropy.org} \citep[version 4.2;][]{AstropyCollaboration2018} and used a more optimized period and duration search grid. 

Using \texttt{Notch}, we detrended the light curve using a 0.3-day filtering window, which removes the stellar variability using a second-order polynomial while preserving the trapezoidal, transit-like signals. We then searched for repeated signals between 0.5 and 30 days with SNR$>$8 using the BLS. We recovered the 6.9 day signal (\planetb) at a BLS SNR of 29 and a 14.3 day signal with a BLS SNR of 17 (\planetname). 

Multiple transit signals of \planetname\ are visible by eye in both the PDCSAP and systematics-corrected light curves before detrending. However, the original implementation of the \notch\ BLS did not recover the second planet signal, nor was it detected when using the PDCSAP flux (independent of the BLS implementation). In addition, using the default SPOC SAP or PDSCAP flux yielded transits with inconsistent depths over the full \tess\ curve. We only recovered the planet when using the updated light curves {\it and} the updated BLS implementation, and only obtained a consistent transit signal when using the updated light curves.

Many of the \planetname\ transits in the \tess\ light curve show evidence of local flares or spot crossings (see Figure \ref{fig:tranist}). The second transit in the Sector 11 data contains such an event during mid-transit, causing an irregular transit shape. These distortions are likely a major contributor to the fact that the planet was not recovered in prior searches. 

The \tess\ SPOC search also recovered a signal from \planetname\ in a multi-sector search of sectors 11, 38 and 64 conducted on 21 June 2023 with a noise-compensating matched filter \citep{Jenkins2020}, although at significantly lower SNR than the notch-based detection. This elevated the detection to a threshold-crossing event (TCE), but it failed the ghost diagnostic test and was never elevated to a \tess\ object of interest (TOI).

\section{Ground-based follow up} \label{sec:groundfollowup}
\subsection{MuSCAT4} \label{sec:muscat}
On 2024 April 22, we observed a predicted transit of \planetname\ using MuSCAT4 \citep[Multicolor Simultaneous Camera for studying Atmospheres of Transiting exoplanets;][]{Narita2020} on the 2m Faulkes Telescope South of Las Cumbres Observatory (LCO) at the Siding Spring observatory in Australia. MuSCAT4 has a field of view of 9\farcm1 $\times$ 9\farcm1 and a pixel scale of 0\farcs27 pixel$^{-1}$. With MuSCAT4, we simultaneously imaged \starname\ in the $g'$, $r'$, $i'$, and $z_s$ bands with exposure times of 12, 7, 8, and 5 seconds, respectively. Due to poor weather, we were unable to observe the transit egress. 

We extracted the light curves using the standard LCOGT BANZAI pipeline \citep{McCully:2018} and applying a customized aperture-photometry pipeline for MuSCAT data \citep{Fukui2011}. We chose an optimal aperture radius and a set of comparison stars for each band that minimized the dispersion of the light curve.

\subsection{Sinistro} \label{sec:lco}
We observed a predicted transit of \planetname\ on 2024 May 6 simultaneously using three LCO 1m telescopes with the Sinistro cameras. For all observations we used the $r'$ filter and an exposure time of 10 seconds.

All the images were calibrated using the standard LCOGT BANZAI pipeline. For two of the three data sets, we performed aperture photometry with \texttt{Photutils} and eight comparison stars. For the other data set, we applied aperture photometry using the same pipeline as used for the MuSCAT4 data. The differences between these extractions were minor, and the resulting three light curves were consistent with each other, so we combined them after applying a small y-offset to ensure the median values matched (a $0.02\%$ correction).

\section{Stellar Properties}

\newedit{\citet{THYMEII} estimated the stellar radius ($R_*=1.38\pm0.06R_\odot$), effective temperature (\teff=$5675\pm75$\,K), and luminosity ($L_*=1.75\pm0.09L_\odot$) of \starname\ by fitting the spectral-energy-distribution with a grid of templates. They estimated the stellar mass ($M_*$) and age by interpolating the observed properties onto the PARSECv1.2S \citep{PARSEC} and BHAC15 \citep{BHAC15} stellar evolutionary models. }

\newedit{Since the discovery, there have been new analyses of the Sco-Cen region that may impact these results. \citet{Ratzenbock2023} found the host star was in the $\nu$-Cen group, which they assign an age between 9.5 and 15.7\,Myr depending on the model and input photometry, consistent with the \citet{THYMEII} estimate (17$\pm$2\,Myr). \citet{SpyglassI} placed \starname{} in the `unclustered' region of Upper Centarus Lupus and were unable to assign a more precise age. }

\newedit{We also repeated the analysis from \citet{THYMEII} using the \gaia{} DR3 parallax and photometry \citep{GaiaCollaboration2023}. All parameters were within 1$\sigma$ of the values from \citet{THYMEII}. Because these and the above age determinations would result in negligible changes, we opted to keep the stellar parameters from \citet{THYMEII} for all analyses.}

\section{Planet properties} \label{sec:transitFit}
\subsection{\tess\ transit fit}
We fit the systematics-corrected \tess\  using \texttt{MISTTBORN} \citep[MCMC Interface for Synthesis of Transits, Tomography, Binaries, and Others of a Relevant Nature;][]{Mann2016a, MISTTBORN}, which uses \texttt{BATMAN} to generate model transits \citep{BATMAN}, \texttt{celerite} to model stellar variability with a Gaussian Process \citep[GP;][]{celerite}, and \texttt{emcee} to explore the parameter space \citep{emcee}. \newedit{Prior to fitting, we manually removed a large flare from 2460059.595 to  2460059.7 BJD.}

We fit both planets (b and c) and the stellar variability simultaneously, with \newedit{19} fit parameters in total. For each planet, we fit for the time of periastron ($T_0$), planet orbital period ($P$), planet-to-star radius ratio ($R_p/R_*$), impact parameter ($b$), and $\sqrt{e}\cos{\omega}$ and $\sqrt{e}\sin{\omega}$ to account for orbital eccentricity ($e$) and the argument of periastron ($\omega$). We fit both planets with a common stellar density ($\rho_*$) and \newedit{quadratic limb-darkening coefficients following triangular sampling} \citep[$q_1$, $q_2$;][]{Kipping2013}.

The remaining \newedit{four} parameters were used to fit for the stellar variability in the GP model. We used a stochastically-driven damped simple harmonic oscillator (SHO)\newedit{, following \cite{Gilbert2022}, with a jitter term}. The free parameters were the \newedit{undampened oscillator period (${P}$), the standard deviation of the process ($\sigma$), the damping timescale ($\tau$), and the jitter term ($\ln{f}$). This is a modified version of the \texttt{RotationTerm} in \texttt{Celerite2}
 \citep{celerite2}.}

Most parameters evolved under uniform priors with only physical limitations. \newedit{The TTV amplitude is small ($<10$\,minutes), with the \tess\ data alone being consistent with a linear ephemeris (Thao et al. submitted), and did not impact our results.} For $\rho_*$, we used a \newedit{Gaussian prior from the SED and isochrone fits reported in \citet{THYMEII}, $\rho_*$/$\rho_\odot$ = $0.46 \pm 0.06$. With this prior on $\rho_*$, the transits primarily constrain eccentricity and $\omega$, instead of $\rho_*$ \citep[see][]{Dawson:2012fk,Van-Eylen2015}. We applied a Gaussian prior on the limb-darkening coefficients ($0.425\pm0.07$ and $0.153\pm0.05$) derived using \texttt{LDTK} toolkit \citep{Parviainen2015} and our adopted stellar parameters.}

\begin{figure*}
    \centering
    \includegraphics[width=.98\linewidth]{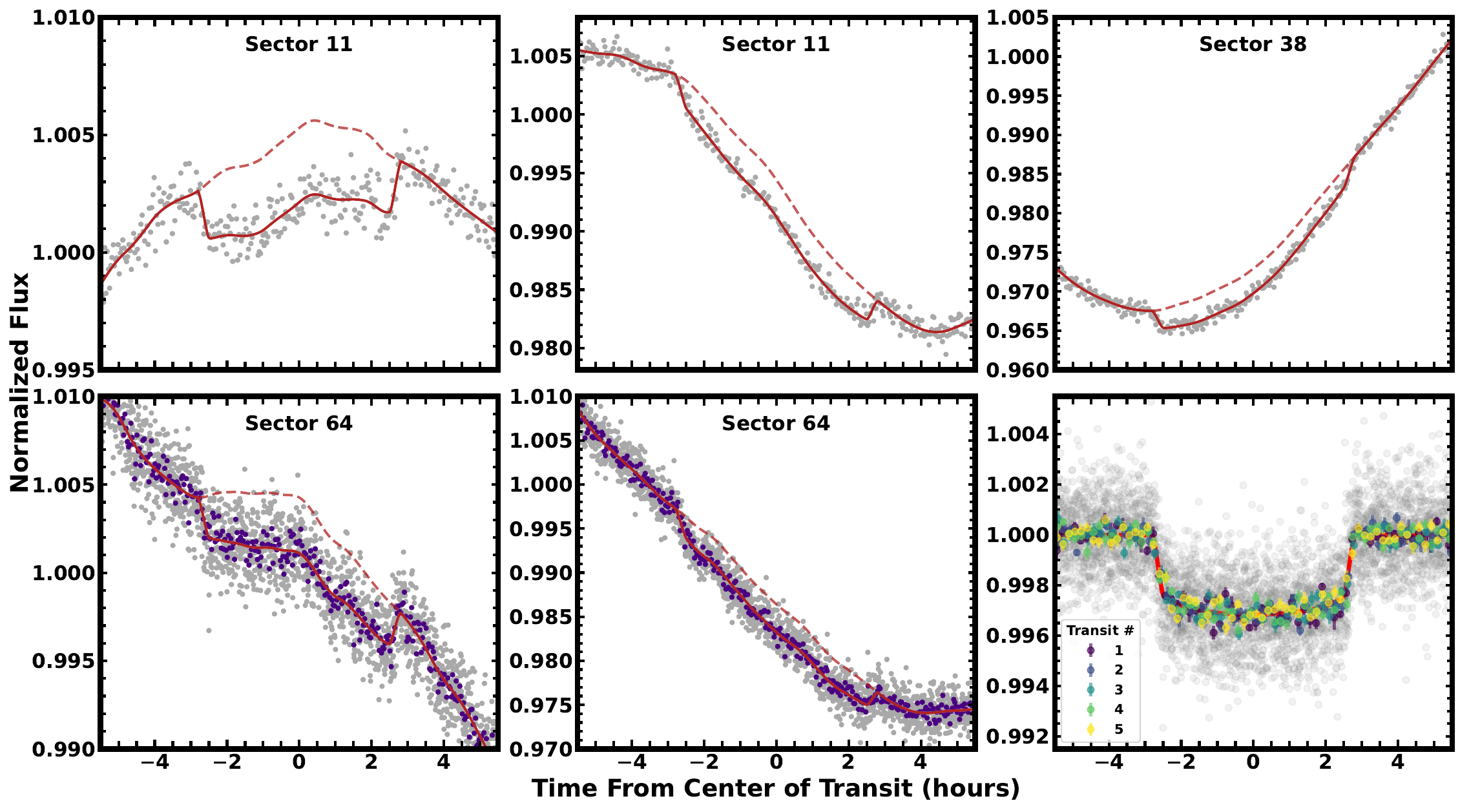}
    \caption{Individual transits of \planetname\ and GP model. Data in Sector 64 (20-second cadence) is binned to 2-minute intervals for easier comparison with Sector 11 and 38 (2-minute cadence). The transits show high levels of in-transit variation (potentially due to spot crossings). Bottom-right: Phase folded light curve of \starname\ taken with \tess\ (gray points) and individual transits binned to 10-minute intervals (colored points) after the GP model of stellar variability has been removed. The best-fit model transit of \planetname\ is shown as the red line.}
    \label{fig:tranist}
\end{figure*}

We ran the MCMC using 50 walkers for 250,000 steps including a 50,000-step burn in. The total run was more than 50 times the autocorrelation time, indicating the number of steps was sufficient for convergence. We present the best-fit parameters in Table \ref{tab:planetParams} and the phase-folded light curve for \planetname\ in Figure \ref{fig:tranist}. We also show a section of the \tess\ light curve with the GP model and locations of transits of both planets in Figure \ref{fig:gpsector}.

\begin{table} 
\scriptsize
\caption{\newedit{Parameters of the \starname\ system from \tess}} 
\hskip-1cm\begin{tabular}{lcc} 
\hline 
\hline
Parameter & \multicolumn{2}{c}{Value} \\ 
\hline
\multicolumn{3}{c}{Stellar Parameters} \\
\hline
$\rho_{\star}$ ($\rho_{\odot}$) & \multicolumn{2}{c}{$0.446^{+0.054}_{-0.047}$} \\ 
$q_{1,1}$ & \multicolumn{2}{c}{$0.153^{+0.049}_{-0.048}$} \\ 
$q_{2,1}$ & \multicolumn{2}{c}{$0.354^{+0.038}_{-0.040}$} \\ 
\hline
\multicolumn{3}{c}{GP Parameters} \\
\hline
$P$ (days) & \multicolumn{2}{c}{$1.177^{+0.033}_{-0.029}$}\\ 
$\sigma$ & \multicolumn{2}{c}{$0.0248^{+0.0013}_{-0.0011}$} \\ 
$\tau$ (days) & \multicolumn{2}{c}{$1.204^{+0.045}_{-0.037}$} \\ 
$\ln{f}$ & \multicolumn{2}{c}{$-13.2^{+1.3}_{-1.2}$} \\ 
\hline
\hline
Parameter & b & c \\
\hline  
\multicolumn{3}{c}{Measured Planet Parameters} \\
\hline 
$T_0$ (BJD-2457000) & $1604.02376^{+0.00033}_{-0.00032}$ & $1602.50256^{+0.00091}_{-0.00093}$\\ 
$P$ (days) & $6.9594731 \pm 2.2\times10^{-6}$ & $14.334892 \pm 1.2\times10^{-5}$\\ 
$R_P/R_{\star}$ & $0.0664^{+0.0015}_{-0.0014}$ & $0.0528^{+0.0023}_{-0.0024}$\\ 
$b$ & $0.03^{+0.19}_{-0.22}$ & $0.26^{+0.20}_{-0.58}$\\ 
$\sqrt{e}\sin\omega$ & $-0.06^{+0.097}_{-0.081}$ & $-0.02^{+0.13}_{-0.11}$\\ 
$\sqrt{e}\cos\omega$ & $-0.07^{+0.32}_{-0.39}$ & $-0.01^{+0.39}_{-0.38}$\\ 
\hline
\multicolumn{3}{c}{Derived Parameters} \\
\hline
$a/R_{\star}$ & $11.66^{+0.24}_{-0.27}$ & $19.14^{+0.63}_{-0.81}$ \\ 
$i$ ($^{\circ}$) & $89.88^{+1.08}_{-0.93}$ & $89.2^{+1.75}_{-0.64}$\\ 
$T_{14}$ (days) & $0.202^{+0.047}_{-0.015}$ & $0.236^{+0.039}_{-0.032}$\\ 
$R_P$ ($R_J$) & $0.891^{+0.021}_{-0.02}$ & $0.708^{+0.031}_{-0.032}$\\ 
$a$ (AU) & $0.0748^{+0.0016}_{-0.0018}$ & $0.1228^{+0.0042}_{-0.0053}$ \\ 
$T_{\mathrm{eq}}$ (K)$^\dagger$ & $1175.0^{+13.0}_{-12.0}$ & $917.0^{+19.0}_{-15.0}$ \\ 
$e$ & $0.064^{+0.187}_{-0.049}$ & $0.077^{+0.195}_{-0.056}$\\ 
$\omega$ ($^{\circ}$) & $195.3^{+140.0}_{-50.0}$ & $186.6^{+160.0}_{-140.0}$\\ 
\hline
\multicolumn{3}{l}{$\dagger$ assuming zero albedo}
\end{tabular}
\label{tab:planetParams}
\end{table}

\subsection{Ground-based transit fits}

\addtolength{\tabcolsep}{-8pt}
\begin{figure*}
\centering
\begin{tabular}{cc}
\raisebox{0cm}{{\includegraphics[width=.48\linewidth]{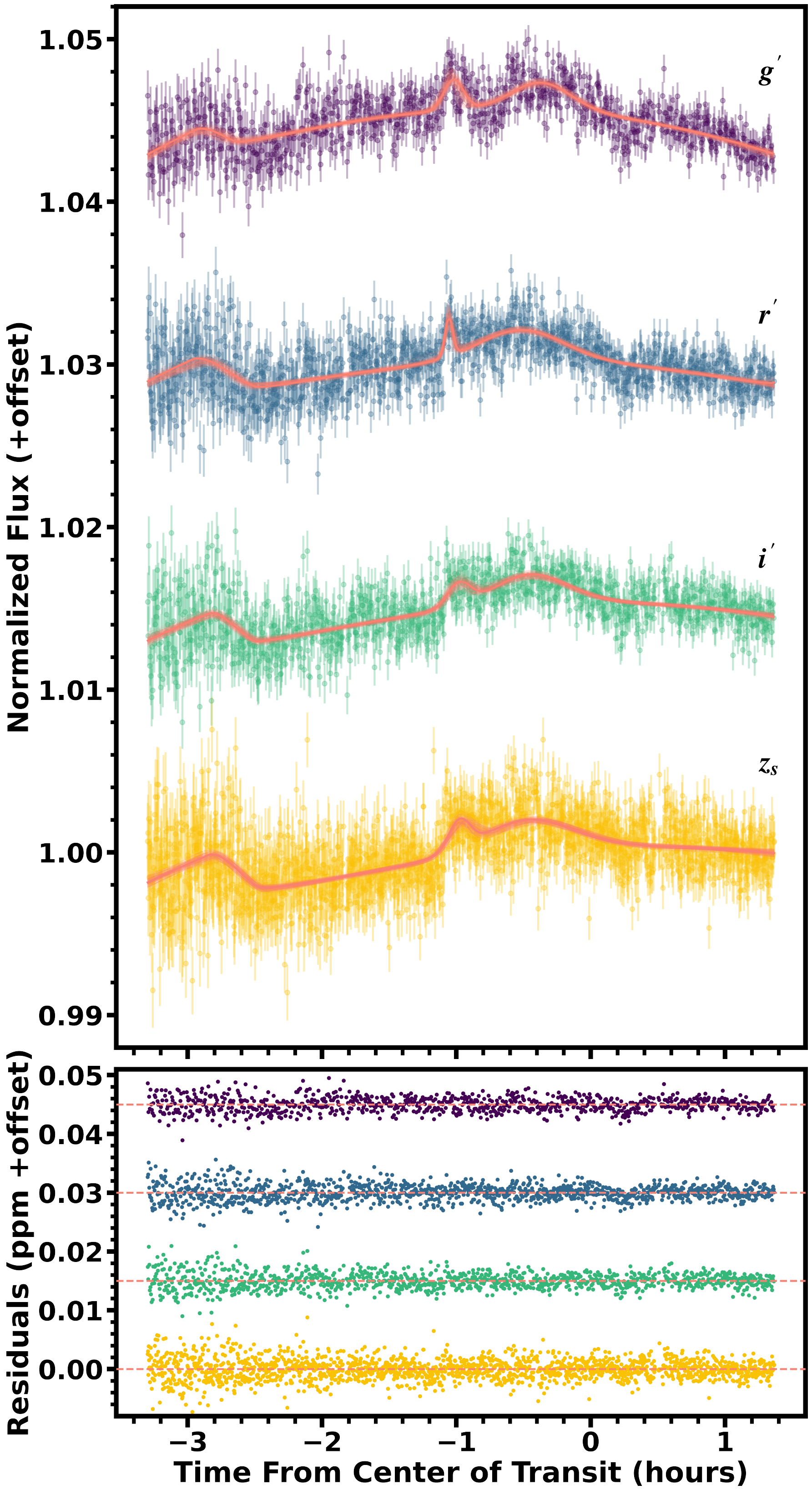}}}
&      
\raisebox{221pt}{\begin{tabular}{c}{%
      \includegraphics[width=.5\linewidth]{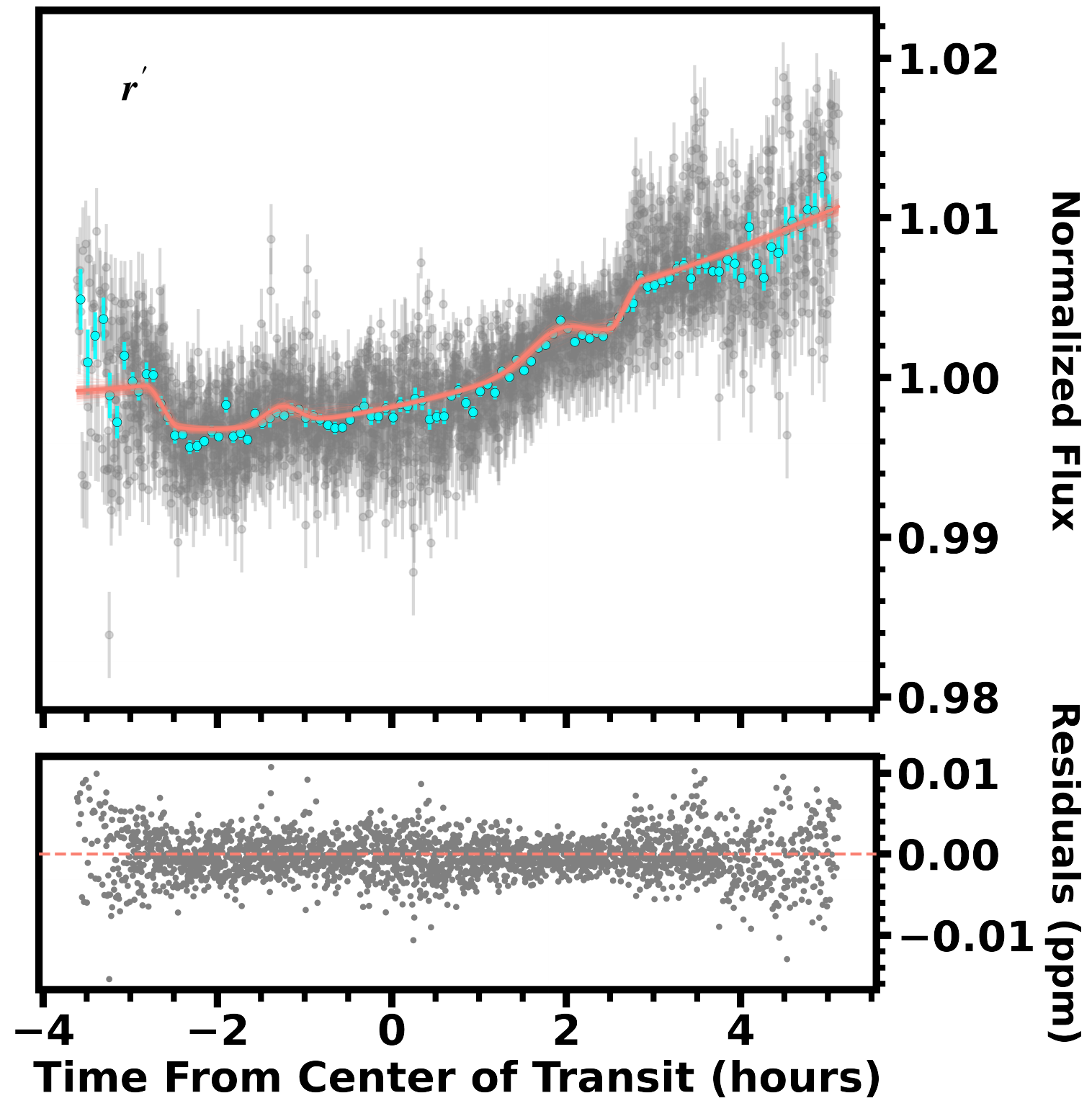}} \\ 
      {\includegraphics[width=.52\linewidth]{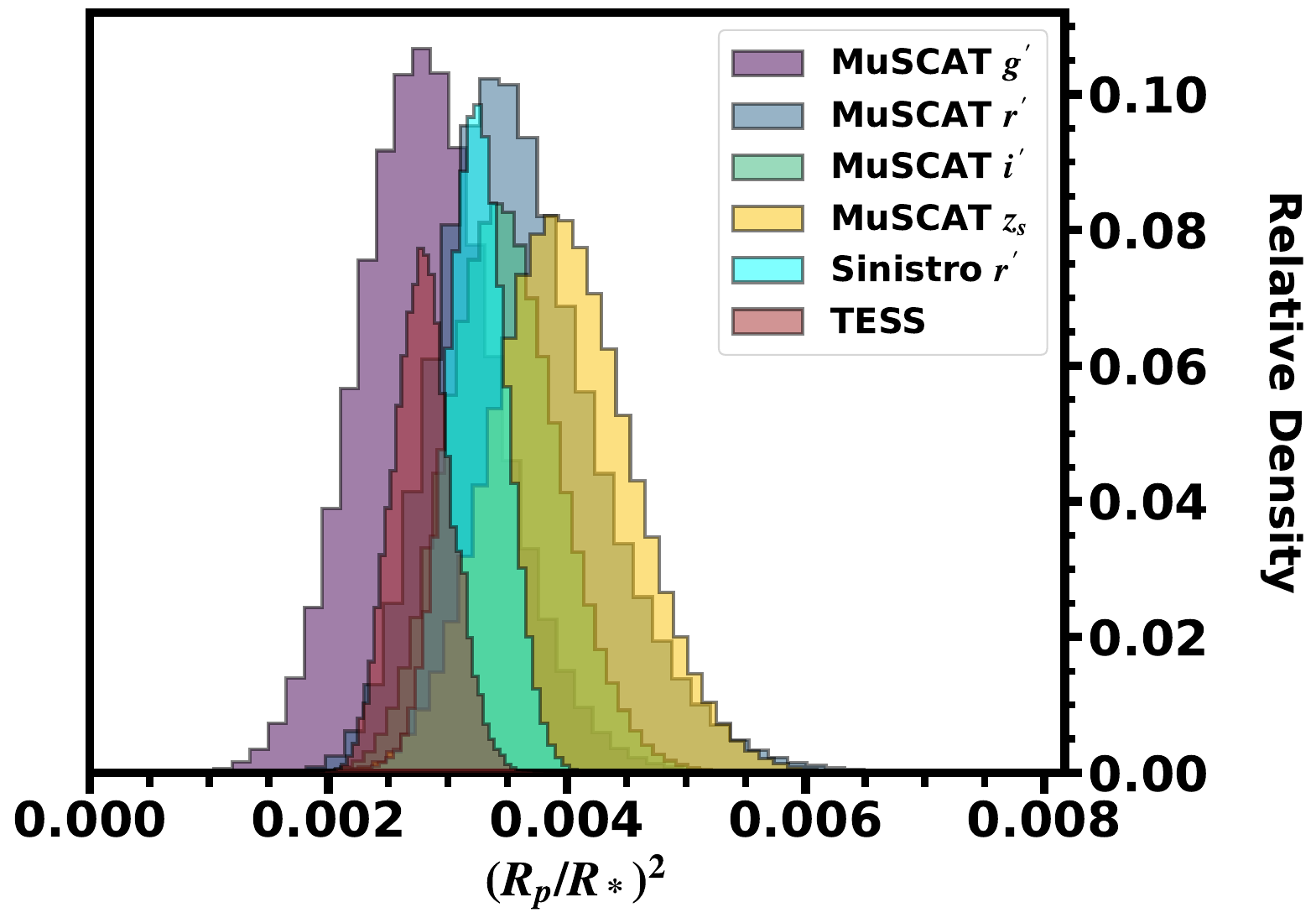}}
\end{tabular}}
\end{tabular}
\caption{Ground-based follow-up transits of \planetname\ from MuSCAT4 (left) and Sinistro (top-right), 5-minute bins shown for clarity. The MuSCAT4 transits were taken simultaneously, with each filter fitted separately. The three simultaneous Sinistro $r'$ transits were stacked and fit together. Each filter was fit with a \texttt{BATMAN} model, two Gaussian spot models, and a second-order polynomial to account for stellar variability. The opaque pink lines are the best-fit model and the translucent lines are 100 models pulled from the fit posterior. The residuals for each transit are also shown. Bottom-right: transit depth, measured by $(R_p/R_*)^2$, posterior from each of the transit fits. }\label{fig:groundTransits}
\end{figure*}
\addtolength{\tabcolsep}{8pt}

Our primary goal for the ground-based data was to confirm the transit depth is consistent with wavelength. To this end, we used an MCMC framework with the \texttt{BATMAN} \citep{BATMAN} transit model. Each fit included a total of 16 parameters.

In the MuSCAT4 data, there is a perturbation between 2460422.985 and 2460423.05 BJD. Similarly, in the Sinistro simultaneous $r'$ transits, all three datasets exhibit a small spike in flux around 2460437.445 BJD. As we discuss further in Section~\ref{sec:concl}, these were most likely spot crossings and were also seen in the \tess\ data. Some events may be flares (particularly the first bump in the MuSCAT4 data), however, our Gaussian spot model (below) described the deviations well. 

To model the spot crossings, we used a Gaussian function, following \cite{Dai2017}. For each spot, we fit for the spot timing ($t_{sp}$), the spot amplitude ($A$), and the spot duration ($\tau$):
\begin{equation}
    Spot(t) = A \times \exp\left[{\frac{-(t - t_{sp})^2}{2\tau^2}}\right].
\end{equation}

We handled the out-of-transit variability using a second-order polynomial, expressed as: 
\begin{equation}
    flux_{corrected} = flux_{raw} \times \left( a + b\times t + c\times t^2 \right).
\end{equation}

We use an additional free parameter ($\ln{f}$) to capture any underestimated uncertainties. The $f$ parameter is an additional fractional uncertainty on the model added in quadrature with the reported uncertainties.

For the transit model, we fit for $R_p/R_*$, $T_0$, semi-major axis-to-star radius ratio ($a/R_*$), orbital inclination ($i$), and two quadratic limb darkening coefficients ($u_1$, $u_2$). For simplicity, we fixed the eccentricity ($e$) to zero. We put a Gaussian prior on the parameters $a/R_*$ and $i$ using the result of the \tess\, fit (Table~\ref{tab:planetParams}) and on $u_{1}$ and $u_{2}$ using the \texttt{LDTK} toolkit \citep{Parviainen2015}. The remaining parameters, $T_0$ and $R_p/R_*$, were allowed to vary within their physically plausible ranges (e.g., $T_0>$ 0 and 0 $< R_{p}/R_{*} <$ 1). 

For each fit, we used 100 walkers for 100,000 steps and a 15,000-step burn in. We fit each of the four bands of the MuSCAT4 data separately, as well as a separate fit to the combined Sinistro $r'$ light curve.

As a test, we modelled the transits with a \texttt{BATMAN} model and a GP as was done for the \tess\ data. We found the fits agreed within uncertainties. Similarly, we tried fitting the transits with the \texttt{BATMAN} model and a second-order polynomial, but opting to mask the spot crossing perturbations instead of explicitly modeling them as Gaussians. Again, all depths agreed within uncertainties. That is, our approach to fitting the data did not change any conclusions with respect to the chomaticity of the transit depth. 

We show each transit fit and residuals, as well as the transit depth posteriors in Figure \ref{fig:groundTransits}. We list the best-fit parameters for each transit in Table \ref{tab:ground}. The depths in each wavelength agreed with each other to $<2\sigma$.

\begin{table*}
\footnotesize
    \centering
    \caption{Ground-based transit fits using Gaussian spot models, a second-order polynomial, and transit model.}
    \begin{tabular}{lccccc}
    \hline 
    \hline
    Parameter & MuSCAT $g'$ & MuSCAT $r'$ & MuSCAT $i'$ & MuSCAT $z_s$ & Sinistro $r'$ \\
    \hline
    \multicolumn{6}{c}{Transit Parameters} \\
    \hline
    $t_{0,O}$ - $t_{0,C}$ (days) & $-0.0049^{+0.0075}_{-0.0098}$ & $-0.0078^{+0.011}_{-0.014}$ & $0.0028^{+0.0074}_{-0.0062}$  & $0.0039^{+0.0076}_{-0.0064}$ & $0.00096^{+0.00087}_{-0.00086}$\\
    $R_p/R_*$ & $0.0531^{+0.0053}_{-0.0054}$ & $0.0595^{+0.0058}_{-0.0050}$ & $0.0588\pm0.0039$ & $0.0629\pm0.0046$ & $0.0569^{+0.0023}_{-0.0021}$\\
    $a/R_*$ & $19.24\pm0.74$ & $19.35\pm0.75$ & $19.21\pm0.73$ & $19.13\pm0.73$ & $19.62^{+0.65}_{-0.58}$\\
    $i$ ($^{\circ}$) & $89.00^{+0.53}_{-0.51}$ & $88.69^{+0.66}_{-0.47}$ & $89.19\pm0.45$ & $89.18^{+0.47}_{-0.46}$ & $89.12^{+0.30}_{-0.36}$\\
    $u_1$ & $0.667\pm0.074$ & $0.471^{+0.075}_{-0.076}$ & $0.378^{+0.074}_{-0.073}$ & $0.370^{+0.076}_{-0.075}$ & $0.321^{+0.070}_{-0.071}$\\
    $u_2$ & $0.047\pm0.040$ & $0.129^{+0.039}_{-0.040}$ & $0.14\pm0.04$ & $0.159^{+0.040}_{-0.039}$ & $0.11\pm0.04$\\
    \hline
    \multicolumn{6}{c}{Stellar Variability Parameters} \\
    \hline
    $a$ & $0.99779\pm0.00019$ & $0.99890\pm0.00020$ & $0.99803\pm0.00018$ & $0.99818^{+0.00022}_{-0.00021}$ & $0.99908^{+0.00024}_{-0.00025}$\\
    $b$ & $0.112^{+0.014}_{-0.013}$ & $0.088^{+0.017}_{-0.014}$ & $0.092^{+0.011}_{-0.012}$ & $0.090\pm0.015$ & $0.0084^{+0.0046}_{-0.0042}$\\
    $c$ & $-0.493^{+0.053}_{-0.059}$ & $-0.36^{+0.06}_{-0.08}$ & $-0.334^{+0.046}_{-0.044}$ & $-0.299^{+0.060}_{-0.058}$ & $0.064^{+0.012}_{-0.013}$\\
    \hline
    \multicolumn{6}{c}{Spot Parameters} \\
    \hline
    $A_1$ & $0.00194^{+0.00034}_{-0.00030}$ & $0.00225^{+0.00046}_{-0.00047}$ & $0.00136^{+0.00027}_{-0.00026}$ & $0.00170\pm0.00028$ & $0.00112^{+0.00027}_{-0.00029}$\\
    $\tau_1$ (days) & $0.00289^{+0.00056}_{-0.00062}$ & $0.00121^{+0.00038}_{-0.00026}$ & $0.00356^{+0.00072}_{-0.00081}$ & $0.00377^{+0.00071}_{-0.00083}$ & $0.0085^{+0.0017}_{-0.0029}$\\
    $t_{sp1}$ (BJD-2457000) & $3422.99112\pm0.00061$ & $3422.98998^{+0.00033}_{-0.00028}$ & $3422.99308\pm0.00094$ & $3422.99358^{+0.00084}_{-0.00092}$ & $3437.3172^{+0.0024}_{-0.0033}$\\
    $A_2$ & $0.00167\pm0.00018$ & $0.00191\pm0.00017$ & $0.00179\pm0.00016$ & $0.00193^{+0.00018}_{-0.00019}$ & $0.00193^{+0.00023}_{-0.00025}$\\
    $\tau_2$ (days) & $0.00932^{+0.00091}_{-0.00083}$ & $0.0138^{+0.0015}_{-0.0013}$ & $0.0119^{+0.0012}_{-0.0013}$ & $0.0142^{+0.0016}_{-0.0019}$ & $0.0142^{+0.0026}_{-0.0033}$\\
    $t_{sp2}$ (BJD-2457000) & $3423.0187\pm0.0010$ & $3423.0119\pm0.0012$ & $3423.0147\pm0.0013$ & $3423.015\pm0.0015$ & $3437.4488^{+0.0018}_{-0.0019}$\\
    $\ln{f}$ & $-7.382^{+0.080}_{-0.089}$ & $-7.55^{+0.10}_{-0.12}$ & $-7.59^{+0.14}_{-0.11}$ & $-7.40^{+0.14}_{-0.11}$ & $-6.513^{+0.033}_{-0.032}$\\
    \hline
    \end{tabular}
    
    \label{tab:ground}
\end{table*}

\section{False positive analysis} \label{sec:falsePositive} 

For an initial assessment, we use \texttt{TRICERATOPS} \citep{triceratops, Giacalone2021}, which calculates the probabilities of various transit-like scenarios in a Bayesian framework. Based on the flattened (GP removed) \tess\ light curves, \texttt{TRICERATOPS} estimated a false-positive probability (FPP) of $< 10^{-6}$. 

While this appears to validate the planet, a statistical false-positive assessment using the light curve morphology, as with \texttt{TRICERATOPS}, may be complicated by spot-crossings and the need to flatten the data. However, an abundance of other evidence strongly indicates \planetname\ is real:
\begin{itemize}[topsep=8pt]
    \item Multi-transiting systems have lower intrinsic (prior) probabilities of being false positives \citep{2012ApJ...750..112L, Rowe2014, Valizadegan2023}.
    \item \newedit{The spot-crossings indicate the transiting/eclipsing body is passing in front of an active star. This favors \starname, as an unassociated background/foreground star is unlikely to be so active.  }
    \item Follow-up imaging, velocities, and color-magnitude diagram position from \cite{THYMEII} already rule out any eclipsing binary or background star. The overwhelming majority of bound companions bright enough to reproduce the c transit would similarly have been detected in the existing follow-up \citep{Wood2021}.
    \item The transit depths are consistent across three instruments spanning more than five years and wavelengths from SDSS $g'$ to $z_s$. This consistency rules out any realistic stellar or instrumental signal. 
    \item The lack of chromaticity sets tight limits on the color of the source of the transit \citep{Desert2015}. Following \citet{THYMEV}, we set a limit of $g'-z_s<2.25$ for the host star. A bound companion within this limit will be $\simeq$20\% as bright as the primary in the optical \citep{PARSEC}. Such a target would be detected in existing high SNR spectra and imaging \citet{Wood2021}, and show as an elevated color-magnitude diagram position \citet{Rizzuto2017}.  
    \item Following \citet{Vanderburg2019}, the transit shape and depth sets the faintest companion which could cause the transit signals to $\Delta T<1.5$\,mags. As with the color constraints, such a star would be detected in one of the suite of follow-up given in \citet{Rizzuto2017} or \citet{Wood2021}.
    \item The detection of TTVs in \planetb\ {\it predicted} the presence of \planetname\ near an integer period ratio. The probability of any false-positive landing near a period ratio by chance is small, and would not explain the TTV seen in \planetb. 
\end{itemize}

We conclude that the signal from \planetname\ is unambiguously a real planet.

\section{Summary and Conclusions} \label{sec:concl}

We report the discovery and validation of \planetname, originally identified as a single transit event by \cite{THYMEII}. We find \planetname\ to be \newedit{0.71 $R_J$ (7.9 $R_\oplus$)} and near the 2:1 mean motion resonance (MMR) with \planetb. We confirm the period and transit depth with two follow-up ground-based transit observations taken with MuSCAT4 and Sinistro. The \starname\ system is now the youngest known transiting multi-planet system to-date.

The \tess\ fit posterior of planet b and c is suggestive of low eccentricity orbits ($<$0.36 and $<$0.39 at 95\% confidence, respectively); the maximum likelihood solution yields an eccentricity of 0.007 for planet b and 0.004 for planet c, with most of the high-eccentricity tail corresponding to specific values of $\omega$ where $e$ variations have low impact on the transit duration. The single transit fit by \cite{THYMEII} found \planetname\ favored a higher eccentricity ($0.29 \pm 0.15$), but this assumed a planetary period $>24$\,days.

Many of the young transiting multi-planet systems are also near MMR (e.g. AU Mic b \& c (9:4) \citep{Plavchan2020, Martioli2021}, V1289 Tau c \& d (3:2) and d \& b (2:1) \citep{David2019a, David2019b, Feinstein2022}, HD 109833 b \& c (3:2) \citep{Wood2023}, and TIC 434398831 b \& c (5:3) \citep{Vach2024, Vach2024b}. Simulations suggest that planets will form in MMR and will begin to drift once the disk dissipates ($\sim$5 Myr) \citep{Izidoro2017}, although many mature planets are found at or near MMR \citep[e.g.][]{Steffen2015}. \citet{Hamer2024} suggest that MMRs should be most common in systems from 10-100\,Myr, and \citet{Dai2024} note that such an excess is visible in the known population of young planets.

The transit depth is consistent across wavelengths from SDSS $g'$ to $z_s$ to within 2$\sigma$, but it is interesting to note the transit depth trends deeper with increasing wavelength. Aside from a statistical coincidence, it may be a bias from the spot crossing and/or additional undetected (and hence not modelled) spot crossings. Alternatively, this may be due to an unocculted hot spot, which will make blue transits appear shallower \citep[e.g.;][]{Rackham2018}.

Our ground-based transits and many of the \tess\ transits show evidence of spot crossings or local flares, far more than is seen in transits of \planetb\ in the \tess\ data. This suggests that \planetname\ crosses a more active latitude of the host star. It may be possible to take advantage of the spot occultations in the \tess\ transits to derive planetary and spot characteristics \citep[e.g,][]{Morris2017} and the stellar obliquity \citep{Desert2011}. Simultaneous multi-band photometry, in particular, can be used to better characterize spot coverage and temperature \citep{Mori2024}.

Unfortunately, our single multi-band transit provides only weak constraints on the spot properties. For the second spot, the amplitude is relatively flat with wavelength, which is more consistent with cool ($<$4500\,K) spots. Although the relatively large uncertainties make it challenging to break the degeneracy between spot size and temperature.

\planetb\ was a Cycle 1 {\it JWST} target \citep{MannJWSTProp}. The resulting transmission spectrum showed strong features, with the transit depth varying by $\simeq$50\% in the CO$_2$ and H$_2$O bands, consistent with a low-density planet (Thao et al. submitted). \planetname\ is only $\simeq$20\% smaller than \planetb\ \newedit{(0.71$R_J$ vs 0.89$R_J$)}. Assuming it also has a low density, as predicted for such young planets \citep{Owen2020}, it will be a similarly high-priority target for \jwst.

\section*{Acknowledgments}
The authors wish to thank Halee, Wally, Penny, and Bandit for their invaluable support. M.G.B. was supported by the NSF Graduate Research Fellowship (DGE-2040435), the NC Space Grant Graduate Research Fellowship, and the TESS Guest Investigator Cycle 5 program (21-TESS21-0016; NASA Grant number 80NSSC24K0880). P.C.T. was supported by the NSF Graduate Research Fellowship (DGE-1650116), the Jack Kent Cooke Foundation Graduate Scholarship, and the JWST GO program (2498). A.W.M. was supported by the NSF CAREER program (AST-2143763) and a grant from NASA's Exoplanet Reseach Program (XRP 80NSSC21K0393). This work is partly supported by JSPS KAKENHI Grant Number JPJP24H00017 and JSPS Bilateral Program Number JPJSBP120249910.

This paper makes use of data collected by the TESS mission. Funding for the TESS mission is provided by NASA’s Science Mission Directorate. We acknowledge the use of public TESS data from pipelines at the TESS Science Office and at the TESS Science Processing Operations Center. Resources supporting this work were provided by the NASA High-End Computing (HEC) Program through the NASA Advanced Supercomputing (NAS) Division at Ames Research Center for the production of the SPOC data products. TESS data presented in this paper were obtained from the Mikulski Archive for Space Telescopes (MAST) at the Space Telescope Science Institute (STScI).

This work makes use of observations from the Las Cumbres Observatory global telescope network. Some data in the paper are based on observations made with the MuSCAT3/4 instruments, developed by the Astrobiology Center (ABC) in Japan, the University of Tokyo, and Las Cumbres Observatory (LCOGT). MuSCAT3 was developed with financial support by JSPS KAKENHI (JP18H05439) and JST PRESTO (JPMJPR1775), and is located at the Faulkes Telescope North on Maui, HI (USA), operated by LCOGT. MuSCAT4 was developed with financial support provided by the Heising-Simons Foundation (grant 2022-3611), JST grant number JPMJCR1761, and the ABC in Japan, and is located at the Faulkes Telescope South at Siding Spring Observatory (Australia), operated by LCOGT.


\vspace{5mm}
\facilities{\tess, LCOGT, MuSCAT}

\software{\texttt{MISTTBORN}, \texttt{astropy}, \texttt{Notch \& LoCoR}, \texttt{TRICERATOPS}}

\bibliography{planetSearch.bib}{}

\begin{thebibliography}{}
\expandafter\ifx\csname natexlab\endcsname\relax\def\natexlab#1{#1}\fi
\providecommand{\url}[1]{\href{#1}{#1}}
\providecommand{\dodoi}[1]{doi:~\href{http://doi.org/#1}{\nolinkurl{#1}}}
\providecommand{\doeprint}[1]{\href{http://ascl.net/#1}{\nolinkurl{http://ascl.net/#1}}}
\providecommand{\doarXiv}[1]{\href{https://arxiv.org/abs/#1}{\nolinkurl{https://arxiv.org/abs/#1}}}

\bibitem[{{Astropy Collaboration} {et~al.}(2018){Astropy Collaboration}, {Price-Whelan}, {Sip{\H{o}}cz}, {G{\"u}nther}, {Lim}, {Crawford}, {Conseil}, {Shupe}, {Craig}, {Dencheva}, {Ginsburg}, {VanderPlas}, {Bradley}, {P{\'e}rez-Su{\'a}rez}, {de Val-Borro}, {Aldcroft}, {Cruz}, {Robitaille}, {Tollerud}, {Ardelean}, {Babej}, {Bach}, {Bachetti}, {Bakanov}, {Bamford}, {Barentsen}, {Barmby}, {Baumbach}, {Berry}, {Biscani}, {Boquien}, {Bostroem}, {Bouma}, {Brammer}, {Bray}, {Breytenbach}, {Buddelmeijer}, {Burke}, {Calderone}, {Cano Rodr{\'\i}guez}, {Cara}, {Cardoso}, {Cheedella}, {Copin}, {Corrales}, {Crichton}, {D'Avella}, {Deil}, {Depagne}, {Dietrich}, {Donath}, {Droettboom}, {Earl}, {Erben}, {Fabbro}, {Ferreira}, {Finethy}, {Fox}, {Garrison}, {Gibbons}, {Goldstein}, {Gommers}, {Greco}, {Greenfield}, {Groener}, {Grollier}, {Hagen}, {Hirst}, {Homeier}, {Horton}, {Hosseinzadeh}, {Hu}, {Hunkeler}, {Ivezi{\'c}}, {Jain}, {Jenness}, {Kanarek}, {Kendrew}, {Kern}, {Kerzendorf}, {Khvalko}, {King}, {Kirkby}, {Kulkarni},
  {Kumar}, {Lee}, {Lenz}, {Littlefair}, {Ma}, {Macleod}, {Mastropietro}, {McCully}, {Montagnac}, {Morris}, {Mueller}, {Mumford}, {Muna}, {Murphy}, {Nelson}, {Nguyen}, {Ninan}, {N{\"o}the}, {Ogaz}, {Oh}, {Parejko}, {Parley}, {Pascual}, {Patil}, {Patil}, {Plunkett}, {Prochaska}, {Rastogi}, {Reddy Janga}, {Sabater}, {Sakurikar}, {Seifert}, {Sherbert}, {Sherwood-Taylor}, {Shih}, {Sick}, {Silbiger}, {Singanamalla}, {Singer}, {Sladen}, {Sooley}, {Sornarajah}, {Streicher}, {Teuben}, {Thomas}, {Tremblay}, {Turner}, {Terr{\'o}n}, {van Kerkwijk}, {de la Vega}, {Watkins}, {Weaver}, {Whitmore}, {Woillez}, {Zabalza}, \& {Astropy Contributors}}]{AstropyCollaboration2018}
{Astropy Collaboration}, {Price-Whelan}, A.~M., {Sip{\H{o}}cz}, B.~M., {et~al.} 2018, \aj, 156, 123, \dodoi{10.3847/1538-3881/aabc4f}

\bibitem[{{Baraffe} {et~al.}(2015){Baraffe}, {Homeier}, {Allard}, \& {Chabrier}}]{BHAC15}
{Baraffe}, I., {Homeier}, D., {Allard}, F., \& {Chabrier}, G. 2015, \aap, 577, A42

\bibitem[{Barat {et~al.}(2024)Barat, Désert, Goyal, Vazan, Kawashima, Fortney, Bean, Line, Panwar, Jacobs, Shivkumar, Sikora, Baeyens, Oklopcić, David, \& Livingston}]{Barat2024}
Barat, S., Désert, J.-M., Goyal, J.~M., {et~al.} 2024, First Comparative Exoplanetology Within a Transiting Multi-planet System: Comparing the atmospheres of V1298 Tau b and c.
\newblock \doarXiv{2407.14995}

\bibitem[{{Bressan} {et~al.}(2012){Bressan}, {Marigo}, {Girardi}, {Salasnich}, {Dal Cero}, {Rubele}, \& {Nanni}}]{PARSEC}
{Bressan}, A., {Marigo}, P., {Girardi}, L., {et~al.} 2012, \mnras, 427, 127, \dodoi{10.1111/j.1365-2966.2012.21948.x}

\bibitem[{{Capistrant} {et~al.}(2024){Capistrant}, {Soares-Furtado}, {Vanderburg}, {Jankowski}, {Mann}, {Ross}, {Srdoc}, {Hinkel}, {Becker}, {Magliano}, {Limbach}, {Stephan}, {Nine}, {Tofflemire}, {Kraus}, {Giacalone}, {Winn}, {Bieryla}, {Bouma}, {Ciardi}, {Collins}, {Covone}, {de Beurs}, {Huang}, {Jenkins}, {Kreidberg}, {Latham}, {Quinn}, {Seager}, {Shporer}, {Twicken}, {Wohler}, {Vanderspek}, {Yarza}, \& {Ziegler}}]{Capistrant2024}
{Capistrant}, B.~K., {Soares-Furtado}, M., {Vanderburg}, A., {et~al.} 2024, \aj, 167, 54, \dodoi{10.3847/1538-3881/ad1039}

\bibitem[{Dai {et~al.}(2017)Dai, Winn, Yu, \& Albrecht}]{Dai2017}
Dai, F., Winn, J.~N., Yu, L., \& Albrecht, S. 2017, The Astronomical Journal, 153, 40

\bibitem[{{Dai} {et~al.}(2024){Dai}, {Goldberg}, {Batygin}, {van Saders}, {Chiang}, {Choksi}, {Li}, {Petigura}, {Gilbert}, {Millholland}, {Dai}, {Bouma}, {Weiss}, \& {Winn}}]{Dai2024}
{Dai}, F., {Goldberg}, M., {Batygin}, K., {et~al.} 2024, arXiv e-prints, arXiv:2406.06885, \dodoi{10.48550/arXiv.2406.06885}

\bibitem[{{David} {et~al.}(2019{\natexlab{a}}){David}, {Petigura}, {Luger}, {Foreman-Mackey}, {Livingston}, {Mamajek}, \& {Hillenbrand}}]{David2019b}
{David}, T.~J., {Petigura}, E.~A., {Luger}, R., {et~al.} 2019{\natexlab{a}}, \apjl, 885, L12, \dodoi{10.3847/2041-8213/ab4c99}

\bibitem[{{David} {et~al.}(2019{\natexlab{b}}){David}, {Cody}, {Hedges}, {Mamajek}, {Hillenbrand}, {Ciardi}, {Beichman}, {Petigura}, {Fulton}, {Isaacson}, {Howard}, {Gagn{\'e}}, {Saunders}, {Rebull}, {Stauffer}, {Vasisht}, \& {Hinkley}}]{David2019a}
{David}, T.~J., {Cody}, A.~M., {Hedges}, C.~L., {et~al.} 2019{\natexlab{b}}, \aj, 158, 79, \dodoi{10.3847/1538-3881/ab290f}

\bibitem[{{Dawson} \& {Johnson}(2012)}]{Dawson:2012fk}
{Dawson}, R.~I., \& {Johnson}, J.~A. 2012, \apj, 756, 122, \dodoi{10.1088/0004-637X/756/2/122}

\bibitem[{{D{\'e}sert} {et~al.}(2011){D{\'e}sert}, {Charbonneau}, {Demory}, {Ballard}, {Carter}, {Fortney}, {Cochran}, {Endl}, {Quinn}, {Isaacson}, {Fressin}, {Buchhave}, {Latham}, {Knutson}, {Bryson}, {Torres}, {Rowe}, {Batalha}, {Borucki}, {Brown}, {Caldwell}, {Christiansen}, {Deming}, {Fabrycky}, {Ford}, {Gilliland}, {Gillon}, {Haas}, {Jenkins}, {Kinemuchi}, {Koch}, {Lissauer}, {Lucas}, {Mullally}, {MacQueen}, {Marcy}, {Sasselov}, {Seager}, {Still}, {Tenenbaum}, {Uddin}, \& {Winn}}]{Desert2011}
{D{\'e}sert}, J.-M., {Charbonneau}, D., {Demory}, B.-O., {et~al.} 2011, \apjs, 197, 14, \dodoi{10.1088/0067-0049/197/1/14}

\bibitem[{{D{\'e}sert} {et~al.}(2015){D{\'e}sert}, {Charbonneau}, {Torres}, {Fressin}, {Ballard}, {Bryson}, {Knutson}, {Batalha}, {Borucki}, {Brown}, {Deming}, {Ford}, {Fortney}, {Gilliland}, {Latham}, \& {Seager}}]{Desert2015}
{D{\'e}sert}, J.-M., {Charbonneau}, D., {Torres}, G., {et~al.} 2015, \apj, 804, 59, \dodoi{10.1088/0004-637X/804/1/59}

\bibitem[{Feinstein {et~al.}(2022)Feinstein, David, Montet, Foreman-Mackey, Livingston, \& Mann}]{Feinstein2022}
Feinstein, A.~D., David, T.~J., Montet, B.~T., {et~al.} 2022, The Astrophysical Journal Letters, 925, l2, \dodoi{10.3847/2041-8213/ac4745}

\bibitem[{{Fernandes} {et~al.}(2022){Fernandes}, {Mulders}, {Pascucci}, {Bergsten}, {Koskinen}, {Hardegree-Ullman}, {Pearson}, {Giacalone}, {Zink}, {Ciardi}, \& {O'Brien}}]{Fernandes2022}
{Fernandes}, R.~B., {Mulders}, G.~D., {Pascucci}, I., {et~al.} 2022, \aj, 164, 78, \dodoi{10.3847/1538-3881/ac7b29}

\bibitem[{{Fernandes} {et~al.}(2023){Fernandes}, {Hardegree-Ullman}, {Pascucci}, {Bergsten}, {Mulders}, {Cunha}, {Mamajek}, {Pearson}, {Feiden}, \& {Curtis}}]{Fernandes2023}
{Fernandes}, R.~B., {Hardegree-Ullman}, K.~K., {Pascucci}, I., {et~al.} 2023, \aj, 166, 175, \dodoi{10.3847/1538-3881/acf4f0}

\bibitem[{{Foreman-Mackey}(2018)}]{celerite2}
{Foreman-Mackey}, D. 2018, Research Notes of the American Astronomical Society, 2, 31, \dodoi{10.3847/2515-5172/aaaf6c}

\bibitem[{{Foreman-Mackey} {et~al.}(2017){Foreman-Mackey}, {Agol}, {Ambikasaran}, \& {Angus}}]{celerite}
{Foreman-Mackey}, D., {Agol}, E., {Ambikasaran}, S., \& {Angus}, R. 2017, \aj, 154, 220, \dodoi{10.3847/1538-3881/aa9332}

\bibitem[{{Foreman-Mackey} {et~al.}(2013){Foreman-Mackey}, {Hogg}, {Lang}, \& {Goodman}}]{emcee}
{Foreman-Mackey}, D., {Hogg}, D.~W., {Lang}, D., \& {Goodman}, J. 2013, \pasp, 125, 306, \dodoi{10.1086/670067}

\bibitem[{{Fukui} {et~al.}(2011){Fukui}, {Narita}, {Tristram}, {Sumi}, {Abe}, {Itow}, {Sullivan}, {Bond}, {Hirano}, {Tamura}, {Bennett}, {Furusawa}, {Hayashi}, {Hearnshaw}, {Hosaka}, {Kamiya}, {Kobara}, {Korpela}, {Kilmartin}, {Lin}, {Ling}, {Makita}, {Masuda}, {Matsubara}, {Miyake}, {Muraki}, {Nagaya}, {Nishimoto}, {Ohnishi}, {Omori}, {Perrott}, {Rattenbury}, {Saito}, {Skuljan}, {Suzuki}, {Sweatman}, \& {Wada}}]{Fukui2011}
{Fukui}, A., {Narita}, N., {Tristram}, P.~J., {et~al.} 2011, \pasj, 63, 287, \dodoi{10.1093/pasj/63.1.287}

\bibitem[{{Gaia Collaboration} {et~al.}(2023){Gaia Collaboration}, {Vallenari}, {Brown}, {Prusti}, {de Bruijne}, {Arenou}, {Babusiaux}, {Biermann}, {Creevey}, {Ducourant}, \& et~al.}]{GaiaCollaboration2023}
{Gaia Collaboration}, {Vallenari}, A., {Brown}, A.~G.~A., {et~al.} 2023, \aap, 674, A1, \dodoi{10.1051/0004-6361/202243940}

\bibitem[{{Giacalone} \& {Dressing}(2020)}]{triceratops}
{Giacalone}, S., \& {Dressing}, C.~D. 2020, {triceratops: Candidate exoplanet rating tool}.
\newblock \doeprint{2002.004}

\bibitem[{{Giacalone} {et~al.}(2021){Giacalone}, {Dressing}, {Jensen}, {Collins}, {Ricker}, {Vanderspek}, {Seager}, {Winn}, {Jenkins}, {Barclay}, {Barkaoui}, {Cadieux}, {Charbonneau}, {Collins}, {Conti}, {Doyon}, {Evans}, {Ghachoui}, {Gillon}, {Guerrero}, {Hart}, {Jehin}, {Kielkopf}, {McLean}, {Murgas}, {Palle}, {Parviainen}, {Pozuelos}, {Relles}, {Shporer}, {Socia}, {Stockdale}, {Tan}, {Torres}, {Twicken}, {Waalkes}, \& {Waite}}]{Giacalone2021}
{Giacalone}, S., {Dressing}, C.~D., {Jensen}, E. L.~N., {et~al.} 2021, \aj, 161, 24, \dodoi{10.3847/1538-3881/abc6af}

\bibitem[{Gilbert {et~al.}(2022)Gilbert, Barclay, Quintana, Walkowicz, Vega, Schlieder, Monsue, Cale, Collins, Gaidos, Mufti, Reefe, Plavchan, Tanner, Wittenmyer, Wittrock, Jenkins, Latham, Ricker, Rose, Seager, Vanderspek, \& Winn}]{Gilbert2022}
Gilbert, E.~A., Barclay, T., Quintana, E.~V., {et~al.} 2022, The Astronomical Journal, 163, 147, \dodoi{10.3847/1538-3881/ac23ca}

\bibitem[{{Hamer} \& {Schlaufman}(2024)}]{Hamer2024}
{Hamer}, J.~H., \& {Schlaufman}, K.~C. 2024, \aj, 167, 55, \dodoi{10.3847/1538-3881/ad110e}

\bibitem[{{Heitzmann} {et~al.}(2021){Heitzmann}, {Zhou}, {Quinn}, {Marsden}, {Wright}, {Petit}, {Vanderburg}, {Bouma}, {Mann}, \& {Rizzuto}}]{Heitzmann2021}
{Heitzmann}, A., {Zhou}, G., {Quinn}, S.~N., {et~al.} 2021, \apjl, 922, L1, \dodoi{10.3847/2041-8213/ac3485}

\bibitem[{Hirano {et~al.}(2020)Hirano, Krishnamurthy, Gaidos, Flewelling, Mann, Narita, Plavchan, Kotani, Tamura, Harakawa, Hodapp, Ishizuka, Jacobson, Konishi, Kudo, Kurokawa, Kuzuhara, Nishikawa, Omiya, Serizawa, Ueda, \& Vievard}]{Hirano2020}
Hirano, T., Krishnamurthy, V., Gaidos, E., {et~al.} 2020, The Astrophysical Journal Letters, 899, L13, \dodoi{10.3847/2041-8213/aba6eb}

\bibitem[{Hirano {et~al.}(2024)Hirano, Gaidos, Harakawa, Hodapp, Kotani, Kudo, Kurokawa, Kuzuhara, Mann, Nishikawa, Omiya, Serizawa, Tamura, Thao, Ueda, \& Vievard}]{Hirano2024}
Hirano, T., Gaidos, E., Harakawa, H., {et~al.} 2024, Monthly Notices of the Royal Astronomical Society, 530, 3117, \dodoi{10.1093/mnras/stae998}

\bibitem[{Howell {et~al.}(2014)Howell, Sobeck, Haas, Still, Barclay, Mullally, Troeltzsch, Aigrain, Bryson, Caldwell, Chaplin, Cochran, Huber, Marcy, Miglio, Najita, Smith, Twicken, \& Fortney}]{Howell2014}
Howell, S.~B., Sobeck, C., Haas, M., {et~al.} 2014, Publications of the Astronomical Society of the Pacific, 126, 398, \dodoi{10.1086/676406}

\bibitem[{{Izidoro} {et~al.}(2017){Izidoro}, {Ogihara}, {Raymond}, {Morbidelli}, {Pierens}, {Bitsch}, {Cossou}, \& {Hersant}}]{Izidoro2017}
{Izidoro}, A., {Ogihara}, M., {Raymond}, S.~N., {et~al.} 2017, \mnras, 470, 1750, \dodoi{10.1093/mnras/stx1232}

\bibitem[{{Jenkins} {et~al.}(2020){Jenkins}, {Tenenbaum}, {Seader}, {Burke}, {McCauliff}, {Smith}, {Twicken}, \& {Chandrasekaran}}]{Jenkins2020}
{Jenkins}, J.~M., {Tenenbaum}, P., {Seader}, S., {et~al.} 2020, {Kepler Data Processing Handbook: Transiting Planet Search}, Kepler Science Document KSCI-19081-003

\bibitem[{{Jenkins} {et~al.}(2016){Jenkins}, {Twicken}, {McCauliff}, {Campbell}, {Sanderfer}, {Lung}, {Mansouri-Samani}, {Girouard}, {Tenenbaum}, {Klaus}, {Smith}, {Caldwell}, {Chacon}, {Henze}, {Heiges}, {Latham}, {Morgan}, {Swade}, {Rinehart}, \& {Vanderspek}}]{Jenkins2016SPOC}
{Jenkins}, J.~M., {Twicken}, J.~D., {McCauliff}, S., {et~al.} 2016, in Society of Photo-Optical Instrumentation Engineers (SPIE) Conference Series, Vol. 9913, \procspie, 99133E, \dodoi{10.1117/12.2233418}

\bibitem[{{Johnson} {et~al.}(2018){Johnson}, {Dai}, {Justesen}, {Gandolfi}, {Hatzes}, {Nowak}, {Endl}, {Cochran}, {Hidalgo}, {Watanabe}, {Parviainen}, {Hirano}, {Villanueva}, {Prieto-Arranz}, {Narita}, {Palle}, {Guenther}, {Barrag{\'a}n}, {Trifonov}, {Niraula}, {MacQueen}, {Cabrera}, {Csizmadia}, {Eigm{\"u}ller}, {Grziwa}, {Korth}, {P{\"a}tzold}, {Smith}, {Albrecht}, {Alonso}, {Deeg}, {Erikson}, {Esposito}, {Fridlund}, {Fukui}, {Kusakabe}, {Kuzuhara}, {Livingston}, {Monta{\~n}es Rodriguez}, {Nespral}, {Persson}, {Purismo}, {Raimundo}, {Rauer}, {Ribas}, {Tamura}, {Van Eylen}, \& {Winn}}]{MISTTBORN}
{Johnson}, M.~C., {Dai}, F., {Justesen}, A.~B., {et~al.} 2018, \mnras, 481, 596, \dodoi{10.1093/mnras/sty2238}

\bibitem[{{Johnson} {et~al.}(2022){Johnson}, {David}, {Petigura}, {Isaacson}, {Van Zandt}, {Ilyin}, {Strassmeier}, {Mallonn}, {Zhou}, {Mann}, {Livingston}, {Luger}, {Dai}, {Weiss}, {Mo{\v{c}}nik}, {Giacalone}, {Hill}, {Rice}, {Blunt}, {Rubenzahl}, {Dalba}, {Esquerdo}, {Berlind}, {Calkins}, \& {Foreman-Mackey}}]{Johnson2022}
{Johnson}, M.~C., {David}, T.~J., {Petigura}, E.~A., {et~al.} 2022, \aj, 163, 247, \dodoi{10.3847/1538-3881/ac6271}

\bibitem[{{Kerr} {et~al.}(2021){Kerr}, {Rizzuto}, {Kraus}, \& {Offner}}]{SpyglassI}
{Kerr}, R., {Rizzuto}, A.~C., {Kraus}, A.~L., \& {Offner}, S. S.~R. 2021, arXiv e-prints, arXiv:2105.09338

\bibitem[{{Kipping}(2013)}]{Kipping2013}
{Kipping}, D.~M. 2013, \mnras, 435, 2152, \dodoi{10.1093/mnras/stt1435}

\bibitem[{{Kreidberg}(2015)}]{BATMAN}
{Kreidberg}, L. 2015, \pasp, 127, 1161, \dodoi{10.1086/683602}

\bibitem[{{Lammers} {et~al.}(2023){Lammers}, {Hadden}, \& {Murray}}]{Lammers2023}
{Lammers}, C., {Hadden}, S., \& {Murray}, N. 2023, \mnras, 525, L66, \dodoi{10.1093/mnrasl/slad092}

\bibitem[{{Lightkurve Collaboration} {et~al.}(2018){Lightkurve Collaboration}, {Cardoso}, {Hedges}, {Gully-Santiago}, {Saunders}, {Cody}, {Barclay}, {Hall}, {Sagear}, {Turtelboom}, {Zhang}, {Tzanidakis}, {Mighell}, {Coughlin}, {Bell}, {Berta-Thompson}, {Williams}, {Dotson}, \& {Barentsen}}]{lightkurve}
{Lightkurve Collaboration}, {Cardoso}, J.~V.~d.~M., {Hedges}, C., {et~al.} 2018, {Lightkurve: Kepler and TESS time series analysis in Python}, Astrophysics Source Code Library.
\newblock \doeprint{1812.013}

\bibitem[{{Lissauer} {et~al.}(2011){Lissauer}, {Ragozzine}, {Fabrycky}, {Steffen}, {Ford}, {Jenkins}, {Shporer}, {Holman}, {Rowe}, {Quintana}, {Batalha}, {Borucki}, {Bryson}, {Caldwell}, {Carter}, {Ciardi}, {Dunham}, {Fortney}, {Gautier}, {Howell}, {Koch}, {Latham}, {Marcy}, {Morehead}, \& {Sasselov}}]{Lissauer2011}
{Lissauer}, J.~J., {Ragozzine}, D., {Fabrycky}, D.~C., {et~al.} 2011, \apjs, 197, 8, \dodoi{10.1088/0067-0049/197/1/8}

\bibitem[{{Lissauer} {et~al.}(2012){Lissauer}, {Marcy}, {Rowe}, {Bryson}, {Adams}, {Buchhave}, {Ciardi}, {Cochran}, {Fabrycky}, {Ford}, {Fressin}, {Geary}, {Gilliland}, {Holman}, {Howell}, {Jenkins}, {Kinemuchi}, {Koch}, {Morehead}, {Ragozzine}, {Seader}, {Tanenbaum}, {Torres}, \& {Twicken}}]{2012ApJ...750..112L}
{Lissauer}, J.~J., {Marcy}, G.~W., {Rowe}, J.~F., {et~al.} 2012, \apj, 750, 112, \dodoi{10.1088/0004-637X/750/2/112}

\bibitem[{{Livingston} {et~al.}(2018){Livingston}, {Dai}, {Hirano}, {Gandolfi}, {Nowak}, {Endl}, {Velasco}, {Fukui}, {Narita}, {Prieto-Arranz}, {Barragan}, {Cusano}, {Albrecht}, {Cabrera}, {Cochran}, {Csizmadia}, {Deeg}, {Eigm{\"u}ller}, {Erikson}, {Fridlund}, {Grziwa}, {Guenther}, {Hatzes}, {Kawauchi}, {Korth}, {Nespral}, {Palle}, {P{\"a}tzold}, {Persson}, {Rauer}, {Smith}, {Tamura}, {Tanaka}, {Van Eylen}, {Watanabe}, \& {Winn}}]{Livingston2018}
{Livingston}, J.~H., {Dai}, F., {Hirano}, T., {et~al.} 2018, \aj, 155, 115, \dodoi{10.3847/1538-3881/aaa841}

\bibitem[{{Mann} {et~al.}(2021){Mann}, {Gao}, {Kraus}, {Newton}, {Thao}, \& {Vanderburg}}]{MannJWSTProp}
{Mann}, A.~W., {Gao}, P., {Kraus}, A.~L., {et~al.} 2021, {The Atmosphere of a 17Myr Old Hot Jupiter}, JWST Proposal. Cycle 1

\bibitem[{{Mann} {et~al.}(2016){Mann}, {Gaidos}, {Mace}, {Johnson}, {Bowler}, {LaCourse}, {Jacobs}, {Vanderburg}, {Kraus}, {Kaplan}, \& {Jaffe}}]{Mann2016a}
{Mann}, A.~W., {Gaidos}, E., {Mace}, G.~N., {et~al.} 2016, \apj, 818, 46, \dodoi{10.3847/0004-637X/818/1/46}

\bibitem[{{Mann} {et~al.}(2022){Mann}, {Wood}, {Schmidt}, {Barber}, {Owen}, {Tofflemire}, {Newton}, {Mamajek}, {Bush}, {Mace}, {Kraus}, {Thao}, {Vanderburg}, {Llama}, {Johns-Krull}, {Prato}, {Stahl}, {Tang}, {Fields}, {Collins}, {Collins}, {Gan}, {Jensen}, {Kamler}, {Schwarz}, {Furlan}, {Gnilka}, {Howell}, {Lester}, {Owens}, {Suarez}, {Mekarnia}, {Guillot}, {Abe}, {Triaud}, {Johnson}, {Milburn}, {Rizzuto}, {Quinn}, {Kerr}, {Ricker}, {Vanderspek}, {Latham}, {Seager}, {Winn}, {Jenkins}, {Guerrero}, {Shporer}, {Schlieder}, {McLean}, \& {Wohler}}]{THYMEVI}
{Mann}, A.~W., {Wood}, M.~L., {Schmidt}, S.~P., {et~al.} 2022, \aj, 163, 156, \dodoi{10.3847/1538-3881/ac511d}

\bibitem[{{Martioli} {et~al.}(2021){Martioli}, {H{\'e}brard}, {Correia}, {Laskar}, \& {Lecavelier des Etangs}}]{Martioli2021}
{Martioli}, E., {H{\'e}brard}, G., {Correia}, A.~C.~M., {Laskar}, J., \& {Lecavelier des Etangs}, A. 2021, \aap, 649, A177, \dodoi{10.1051/0004-6361/202040235}

\bibitem[{{Masuda}(2014)}]{Masuda2014}
{Masuda}, K. 2014, \apj, 783, 53, \dodoi{10.1088/0004-637X/783/1/53}

\bibitem[{{McCully} {et~al.}(2018){McCully}, {Volgenau}, {Harbeck}, {Lister}, {Saunders}, {Turner}, {Siiverd}, \& {Bowman}}]{McCully:2018}
{McCully}, C., {Volgenau}, N.~H., {Harbeck}, D.-R., {et~al.} 2018, in Society of Photo-Optical Instrumentation Engineers (SPIE) Conference Series, Vol. 10707, \procspie, 107070K, \dodoi{10.1117/12.2314340}

\bibitem[{Mori {et~al.}(2024)Mori, Ikuta, Fukui, Narita, de Leon, Livingston, Ikoma, Kawai, Kawauchi, Murgas, Palle, Parviainen, Rodríguez, Terada, Watanabe, \& Tamura}]{Mori2024}
Mori, M., Ikuta, K., Fukui, A., {et~al.} 2024, Monthly Notices of the Royal Astronomical Society, 530, 167, \dodoi{10.1093/mnras/stae841}

\bibitem[{Morris {et~al.}(2017)Morris, Hebb, Davenport, Rohn, \& Hawley}]{Morris2017}
Morris, B.~M., Hebb, L., Davenport, J. R.~A., Rohn, G., \& Hawley, S.~L. 2017, The Astrophysical Journal, 846, 99, \dodoi{10.3847/1538-4357/aa8555}

\bibitem[{{Morris} {et~al.}(2020){Morris}, {Twicken}, {Smith}, {Clarke}, {Jenkins}, {Bryson}, {Girouard}, \& {Klaus}}]{Morris2020SPOC}
{Morris}, R.~L., {Twicken}, J.~D., {Smith}, J.~C., {et~al.} 2020, {Kepler Data Processing Handbook: Photometric Analysis}, Kepler Science Document KSCI-19081-003, id. 6. Edited by Jon M. Jenkins.

\bibitem[{{Nardiello} {et~al.}(2022){Nardiello}, {Malavolta}, {Desidera}, {Baratella}, {D'Orazi}, {Messina}, {Biazzo}, {Benatti}, {Damasso}, {Rajpaul}, {Bonomo}, {Capuzzo Dolcetta}, {Mallonn}, {Cale}, {Plavchan}, {El Mufti}, {Bignamini}, {Borsa}, {Carleo}, {Claudi}, {Covino}, {Lanza}, {Maldonado}, {Mancini}, {Micela}, {Molinari}, {Pinamonti}, {Piotto}, {Poretti}, {Scandariato}, {Sozzetti}, {Andreuzzi}, {Boschin}, {Cosentino}, {Fiorenzano}, {Harutyunyan}, {Knapic}, {Pedani}, {Affer}, {Maggio}, \& {Rainer}}]{Nardiello2022}
{Nardiello}, D., {Malavolta}, L., {Desidera}, S., {et~al.} 2022, \aap, 664, A163, \dodoi{10.1051/0004-6361/202243743}

\bibitem[{{Narita} {et~al.}(2020){Narita}, {Fukui}, {Yamamuro}, {Harbeck}, {Bowman}, {Elphick}, {Nation}, {Armstrong}, {Han}, {Abe}, {Ikoma}, {Isogai}, {Kawauchi}, {Kurita}, {Kusakabe}, {de Leon}, {Livingston}, {Mori}, {Nishiumi}, {Tamura}, {Watanabe}, {Volgenau}, {Heinrich-Josties}, {Foale}, {Daily}, {McCully}, {Kirby}, {Smith}, {Haworth}, {Conway}, {Storrie-Lombardi}, {Rosing}, {Chatelain}, {Bachelet}, {Johnson}, \& {Rabus}}]{Narita2020}
{Narita}, N., {Fukui}, A., {Yamamuro}, T., {et~al.} 2020, in Society of Photo-Optical Instrumentation Engineers (SPIE) Conference Series, Vol. 11447, Ground-based and Airborne Instrumentation for Astronomy VIII, ed. C.~J. {Evans}, J.~J. {Bryant}, \& K.~{Motohara}, 114475K, \dodoi{10.1117/12.2559947}

\bibitem[{{Owen}(2020)}]{Owen2020}
{Owen}, J.~E. 2020, \mnras, 498, 5030, \dodoi{10.1093/mnras/staa2784}

\bibitem[{{Parviainen} \& {Aigrain}(2015)}]{Parviainen2015}
{Parviainen}, H., \& {Aigrain}, S. 2015, \mnras, 453, 3821, \dodoi{10.1093/mnras/stv1857}

\bibitem[{{Plavchan} {et~al.}(2020){Plavchan}, {Barclay}, {Gagn{\'e}}, {Gao}, {Cale}, {Matzko}, {Dragomir}, {Quinn}, {Feliz}, {Stassun}, {Crossfield}, {Berardo}, {Latham}, {Tieu}, {Anglada-Escud{\'e}}, {Ricker}, {Vanderspek}, {Seager}, {Winn}, {Jenkins}, {Rinehart}, {Krishnamurthy}, {Dynes}, {Doty}, {Adams}, {Afanasev}, {Beichman}, {Bottom}, {Bowler}, {Brinkworth}, {Brown}, {Cancino}, {Ciardi}, {Clampin}, {Clark}, {Collins}, {Davison}, {Foreman-Mackey}, {Furlan}, {Gaidos}, {Geneser}, {Giddens}, {Gilbert}, {Hall}, {Hellier}, {Henry}, {Horner}, {Howard}, {Huang}, {Huber}, {Kane}, {Kenworthy}, {Kielkopf}, {Kipping}, {Klenke}, {Kruse}, {Latouf}, {Lowrance}, {Mennesson}, {Mengel}, {Mills}, {Morton}, {Narita}, {Newton}, {Nishimoto}, {Okumura}, {Palle}, {Pepper}, {Quintana}, {Roberge}, {Roccatagliata}, {Schlieder}, {Tanner}, {Teske}, {Tinney}, {Vanderburg}, {von Braun}, {Walp}, {Wang}, {Wang}, {Weigand}, {White}, {Wittenmyer}, {Wright}, {Youngblood}, {Zhang}, \& {Zilberman}}]{Plavchan2020}
{Plavchan}, P., {Barclay}, T., {Gagn{\'e}}, J., {et~al.} 2020, \nat, 582, 497, \dodoi{10.1038/s41586-020-2400-z}

\bibitem[{{Rackham} {et~al.}(2018){Rackham}, {Apai}, \& {Giampapa}}]{Rackham2018}
{Rackham}, B.~V., {Apai}, D., \& {Giampapa}, M.~S. 2018, \apj, 853, 122, \dodoi{10.3847/1538-4357/aaa08c}

\bibitem[{{Ratzenb{\"o}ck} {et~al.}(2023){Ratzenb{\"o}ck}, {Gro{\ss}schedl}, {M{\"o}ller}, {Alves}, {Bomze}, \& {Meingast}}]{Ratzenbock2023}
{Ratzenb{\"o}ck}, S., {Gro{\ss}schedl}, J.~E., {M{\"o}ller}, T., {et~al.} 2023, \aap, 677, A59

\bibitem[{{Ricker} {et~al.}(2015){Ricker}, {Winn}, {Vanderspek}, {Latham}, {Bakos}, {Bean}, {Berta-Thompson}, {Brown}, {Buchhave}, {Butler}, {Butler}, {Chaplin}, {Charbonneau}, {Christensen-Dalsgaard}, {Clampin}, {Deming}, {Doty}, {De Lee}, {Dressing}, {Dunham}, {Endl}, {Fressin}, {Ge}, {Henning}, {Holman}, {Howard}, {Ida}, {Jenkins}, {Jernigan}, {Johnson}, {Kaltenegger}, {Kawai}, {Kjeldsen}, {Laughlin}, {Levine}, {Lin}, {Lissauer}, {MacQueen}, {Marcy}, {McCullough}, {Morton}, {Narita}, {Paegert}, {Palle}, {Pepe}, {Pepper}, {Quirrenbach}, {Rinehart}, {Sasselov}, {Sato}, {Seager}, {Sozzetti}, {Stassun}, {Sullivan}, {Szentgyorgyi}, {Torres}, {Udry}, \& {Villasenor}}]{Ricker2015}
{Ricker}, G.~R., {Winn}, J.~N., {Vanderspek}, R., {et~al.} 2015, Journal of Astronomical Telescopes, Instruments, and Systems, 1, 014003, \dodoi{10.1117/1.JATIS.1.1.014003}

\bibitem[{{Rizzuto} {et~al.}(2017){Rizzuto}, {Mann}, {Vanderburg}, {Kraus}, \& {Covey}}]{Rizzuto2017}
{Rizzuto}, A.~C., {Mann}, A.~W., {Vanderburg}, A., {Kraus}, A.~L., \& {Covey}, K.~R. 2017, \aj, 154, 224, \dodoi{10.3847/1538-3881/aa9070}

\bibitem[{{Rizzuto} {et~al.}(2020){Rizzuto}, {Newton}, {Mann}, {Tofflemire}, {Vanderburg}, {Kraus}, {Wood}, {Quinn}, {Zhou}, {Thao}, {Law}, {Ziegler}, \& {Brice{\~n}o}}]{THYMEII}
{Rizzuto}, A.~C., {Newton}, E.~R., {Mann}, A.~W., {et~al.} 2020, \aj, 160, 33, \dodoi{10.3847/1538-3881/ab94b7}

\bibitem[{{Rowe} {et~al.}(2014){Rowe}, {Bryson}, {Marcy}, {Lissauer}, {Jontof-Hutter}, {Mullally}, {Gilliland}, {Issacson}, {Ford}, {Howell}, {Borucki}, {Haas}, {Huber}, {Steffen}, {Thompson}, {Quintana}, {Barclay}, {Still}, {Fortney}, {Gautier}, {Hunter}, {Caldwell}, {Ciardi}, {Devore}, {Cochran}, {Jenkins}, {Agol}, {Carter}, \& {Geary}}]{Rowe2014}
{Rowe}, J.~F., {Bryson}, S.~T., {Marcy}, G.~W., {et~al.} 2014, \apj, 784, 45, \dodoi{10.1088/0004-637X/784/1/45}

\bibitem[{{Smith} {et~al.}(2012){Smith}, {Stumpe}, {Van Cleve}, {Jenkins}, {Barclay}, {Fanelli}, {Girouard}, {Kolodziejczak}, {McCauliff}, {Morris}, \& {Twicken}}]{Smith2012}
{Smith}, J.~C., {Stumpe}, M.~C., {Van Cleve}, J.~E., {et~al.} 2012, \pasp, 124, 1000, \dodoi{10.1086/667697}

\bibitem[{{Soderblom} {et~al.}(2014){Soderblom}, {Hillenbrand}, {Jeffries}, {Mamajek}, \& {Naylor}}]{Soderblom2014}
{Soderblom}, D.~R., {Hillenbrand}, L.~A., {Jeffries}, R.~D., {Mamajek}, E.~E., \& {Naylor}, T. 2014, in Protostars and Planets VI, ed. H.~{Beuther}, R.~S. {Klessen}, C.~P. {Dullemond}, \& T.~{Henning}, 219--241, \dodoi{10.2458/azu_uapress_9780816531240-ch010}

\bibitem[{{Steffen} \& {Hwang}(2015)}]{Steffen2015}
{Steffen}, J.~H., \& {Hwang}, J.~A. 2015, \mnras, 448, 1956, \dodoi{10.1093/mnras/stv104}

\bibitem[{{Steffen} {et~al.}(2012){Steffen}, {Fabrycky}, {Ford}, {Carter}, {D{\'e}sert}, {Fressin}, {Holman}, {Lissauer}, {Moorhead}, {Rowe}, {Ragozzine}, {Welsh}, {Batalha}, {Borucki}, {Buchhave}, {Bryson}, {Caldwell}, {Charbonneau}, {Ciardi}, {Cochran}, {Endl}, {Everett}, {Gautier}, {Gilliland}, {Girouard}, {Jenkins}, {Horch}, {Howell}, {Isaacson}, {Klaus}, {Koch}, {Latham}, {Li}, {Lucas}, {MacQueen}, {Marcy}, {McCauliff}, {Middour}, {Morris}, {Mullally}, {Quinn}, {Quintana}, {Shporer}, {Still}, {Tenenbaum}, {Thompson}, {Twicken}, \& {Van Cleve}}]{Steffen2012}
{Steffen}, J.~H., {Fabrycky}, D.~C., {Ford}, E.~B., {et~al.} 2012, \mnras, 421, 2342, \dodoi{10.1111/j.1365-2966.2012.20467.x}

\bibitem[{{Stumpe} {et~al.}(2014){Stumpe}, {Smith}, {Catanzarite}, {Van Cleve}, {Jenkins}, {Twicken}, \& {Girouard}}]{Stumpe2014}
{Stumpe}, M.~C., {Smith}, J.~C., {Catanzarite}, J.~H., {et~al.} 2014, \pasp, 126, 100, \dodoi{10.1086/674989}

\bibitem[{{Stumpe} {et~al.}(2012){Stumpe}, {Smith}, {Van Cleve}, {Twicken}, {Barclay}, {Fanelli}, {Girouard}, {Jenkins}, {Kolodziejczak}, {McCauliff}, \& {Morris}}]{Stumpe2012}
{Stumpe}, M.~C., {Smith}, J.~C., {Van Cleve}, J.~E., {et~al.} 2012, \pasp, 124, 985, \dodoi{10.1086/667698}

\bibitem[{Thao {et~al.}(2024)Thao, Mann, Barber, Kraus, Tofflemire, Bush, Wood, Collins, Vanderburg, Quinn, Zhou, Newton, Ziegler, Law, Barkaoui, Pozuelos, Timmermans, Gillon, Jehin, Schwarz, Gan, Shporer, Horne, Sefako, Suarez, Mekarnia, Guillot, Abe, Triaud, Radford, Murillo, Ricker, Winn, Jenkins, Bouma, Fausnaugh, Guerrero, \& Kunimoto}]{Thao2024}
Thao, P.~C., Mann, A.~W., Barber, M.~G., {et~al.} 2024, The Astronomical Journal, 168, 41, \dodoi{10.3847/1538-3881/ad4993}

\bibitem[{{Tofflemire} {et~al.}(2021){Tofflemire}, {Rizzuto}, {Newton}, {Kraus}, {Mann}, {Vanderburg}, {Nelson}, {Hawkins}, {Wood}, {Zhou}, {Quinn}, {Howell}, {Collins}, {Schwarz}, {Stassun}, {Bouma}, {Essack}, {Osborn}, {Boyd}, {F{\H{u}}r{\'e}sz}, {Glidden}, {Twicken}, {Wohler}, {McLean}, {Ricker}, {Vanderspek}, {Latham}, {Seager}, {Winn}, \& {Jenkins}}]{THYMEV}
{Tofflemire}, B.~M., {Rizzuto}, A.~C., {Newton}, E.~R., {et~al.} 2021, \aj, 161, 171, \dodoi{10.3847/1538-3881/abdf53}

\bibitem[{{Twicken} {et~al.}(2010){Twicken}, {Clarke}, {Bryson}, {Tenenbaum}, {Wu}, {Jenkins}, {Girouard}, \& {Klaus}}]{Twicken2010SAP}
{Twicken}, J.~D., {Clarke}, B.~D., {Bryson}, S.~T., {et~al.} 2010, in Society of Photo-Optical Instrumentation Engineers (SPIE) Conference Series, Vol. 7740, Software and Cyberinfrastructure for Astronomy, ed. N.~M. {Radziwill} \& A.~{Bridger}, 774023, \dodoi{10.1117/12.856790}

\bibitem[{{Vach} {et~al.}(2024{\natexlab{a}}){Vach}, {Zhou}, {Huang}, {Rogers}, {Bouma}, {Douglas}, {Kunimoto}, {Mann}, {Barber}, {Quinn}, {Latham}, {Bieryla}, \& {Collins}}]{Vach2024}
{Vach}, S., {Zhou}, G., {Huang}, C.~X., {et~al.} 2024{\natexlab{a}}, \aj, 167, 210, \dodoi{10.3847/1538-3881/ad3108}

\bibitem[{{Vach} {et~al.}(2024{\natexlab{b}}){Vach}, Zhou, Huang, Mann, Barber, Bieryla, Latham, Collins, Rogers, Bouma, Douglas, Quinn, Fairnington, Krüger, Shporer, Collins, Srdoc, Schwarz, Relles, Barkaoui, McLeod, Schneider, Narita, Fukui, Sefako, Fong, Mireles, Torres, Ricker, Seager, \& Winn}]{Vach2024b}
{Vach}, S., Zhou, G., Huang, C.~X., {et~al.} 2024{\natexlab{b}}, A transiting multi-planet system in the 61 million year old association Theia 116.
\newblock \doarXiv{2407.19680}

\bibitem[{{Valizadegan} {et~al.}(2023){Valizadegan}, {Martinho}, {Jenkins}, {Caldwell}, {Twicken}, \& {Bryson}}]{Valizadegan2023}
{Valizadegan}, H., {Martinho}, M. J.~S., {Jenkins}, J.~M., {et~al.} 2023, \aj, 166, 28, \dodoi{10.3847/1538-3881/acd344}

\bibitem[{{Van Eylen} \& {Albrecht}(2015)}]{Van-Eylen2015}
{Van Eylen}, V., \& {Albrecht}, S. 2015, \apj, 808, 126, \dodoi{10.1088/0004-637X/808/2/126}

\bibitem[{{Vanderburg} {et~al.}(2019){Vanderburg}, {Huang}, {Rodriguez}, {Becker}, {Ricker}, {Vanderspek}, {Latham}, {Seager}, {Winn}, {Jenkins}, {Addison}, {Bieryla}, {Brice{\~n}o}, {Bowler}, {Brown}, {Burke}, {Burt}, {Caldwell}, {Clark}, {Crossfield}, {Dittmann}, {Dynes}, {Fulton}, {Guerrero}, {Harbeck}, {Horner}, {Kane}, {Kielkopf}, {Kraus}, {Kreidberg}, {Law}, {Mann}, {Mengel}, {Morton}, {Okumura}, {Pearce}, {Plavchan}, {Quinn}, {Rabus}, {Rose}, {Rowden}, {Shporer}, {Siverd}, {Smith}, {Stassun}, {Tinney}, {Wittenmyer}, {Wright}, {Zhang}, {Zhou}, \& {Ziegler}}]{Vanderburg2019}
{Vanderburg}, A., {Huang}, C.~X., {Rodriguez}, J.~E., {et~al.} 2019, \apjl, 881, L19, \dodoi{10.3847/2041-8213/ab322d}

\bibitem[{Wood {et~al.}(2021)Wood, Mann, \& Kraus}]{Wood2021}
Wood, M.~L., Mann, A.~W., \& Kraus, A.~L. 2021, The Astronomical Journal, 162, 128, \dodoi{10.3847/1538-3881/ac0ae9}

\bibitem[{{Wood} {et~al.}(2023){Wood}, {Mann}, {Barber}, {Bush}, {Kraus}, {Tofflemire}, {Vanderburg}, {Newton}, {Feiden}, {Zhou}, {Bouma}, {Quinn}, {Armstrong}, {Osborn}, {Adibekyan}, {Mena}, {Sousa}, {Gagn{\'e}}, {Fields}, {Milburn}, {Thao}, {Schmidt}, {Gnilka}, {Howell}, {Law}, {Ziegler}, {Brice{\~n}o}, {Ricker}, {Vanderspek}, {Latham}, {Seager}, {Winn}, {Jenkins}, {Schlieder}, {Osborn}, {Twicken}, {Ciardi}, \& {Huang}}]{Wood2023}
{Wood}, M.~L., {Mann}, A.~W., {Barber}, M.~G., {et~al.} 2023, \aj, 165, 85, \dodoi{10.3847/1538-3881/aca8fc}

\bibitem[{{Zakhozhay} {et~al.}(2022){Zakhozhay}, {Launhardt}, {Trifonov}, {K{\"u}rster}, {Reffert}, {Henning}, {Brahm}, {Vin{\'e}s}, {Marleau}, \& {Patel}}]{Zakhozhay2022}
{Zakhozhay}, O.~V., {Launhardt}, R., {Trifonov}, T., {et~al.} 2022, \aap, 667, L14, \dodoi{10.1051/0004-6361/202244747}

\end{thebibliography}
\bibliographystyle{aasjournal}

\end{document}